\documentclass[structabstract]{aa}
\usepackage{txfonts}
\usepackage{graphicx}
\usepackage{natbib}
\usepackage{lscape}
\usepackage{wasysym}
\usepackage{hyperref}
\bibpunct{(}{)}{;}{a}{}{,} % to follow the A&A style

\begin{document}

\title{Planetary detection limits taking into account stellar noise}
%\thanks{Based on observations collected at the La Silla Parana Observatory,
%ESO (Chile), with the HARPS spectrograph at the 3.6-m telescope.}

\subtitle{II. Effect of stellar spot groups on radial-velocities}

\author{
  X. Dumusque\inst{1,2} \and
  N.C. Santos\inst{1,3} \and
  S. Udry\inst{2} \and
  C. Lovis\inst{2} \and
  X. Bonfils\inst{4}
  }

\institute{
    Centro de Astrof{\'\i}sica, Universidade do Porto, Rua das Estrelas, 4150-762 Porto, Portugal \and
    Observatoire de Gen\`eve, Universit\'e de Genve, 51 ch. des Maillettes, 1290 Sauverny, Switzerland \and
    Departamento de F\'isica e Astronomia, Faculdade de Ci\^encias, Universidade do Porto, Portugal \and
    Universit\'e J. Fourier (Grenoble 1)/CNRS, Laboratoire dÕAstrophysique de Grenoble (LAOG, UMR5571), France
}

\date{Received XXX; accepted XXX}

\abstract
{The detection of small mass planets with the radial-velocity technique is now confronted with the interference of stellar noise. HARPS can now reach a precision below the meter-per-second, which corresponds to the amplitudes of different stellar perturbations, such as oscillation, granulation, and activity. }
{Solar spot groups induced by activity produce a radial-velocity noise of a few meter-per-second. The aim of this paper is to simulate this activity and calculate detection limits according to different observational strategies.}
{Based on Sun observations, we reproduce the evolution of spot groups on the surface of a rotating star. We then calculate the radial-velocity effect induced by these spot groups as a function of time. Taking into account oscillation, granulation, activity, and a HARPS instrumental error of 80\,cm\,s$^{-1}$, we simulate the effect of different observational strategies in order to efficiently reduce all sources of noise.}
{Applying three measurements per night of 10 minutes every three days, 10 nights a month seems the best tested strategy. Depending on the level of activity considered, from $\log{R'_{HK}}= -5$ to $-4.75$, this strategy would allow us to find planets of 2.5 to 3.5\,M$_{\oplus}$ in the habitable zone of a K1V dwarf. Using Bern's model of planetary formation, we estimate that for the same range of activity level, 15 to 35\,\% of the planets between 1 and 5\,M$_{\oplus}$ and with a period between 100 and 200 days should be found with HARPS. A comparison between the performance of HARPS and ESPRESSO is also emphasized by our simulations. Using the same optimized strategy, ESPRESSO could find 1.3\,M$_{\oplus}$ planets in the habitable zone of K dwarfs. In addition, 80\,\% of planets with mass between 1 and 5\,M$_{\oplus}$ and with a period between 100 and 200 days could be detected.}
{}

\authorrunning{Dumusque et al.}
\titlerunning{Activity noise and planetary detection}
\maketitle

\section{Introduction}

We have now discovered more than 400 exoplanets with the radial-velocity (RV) technique\footnote{see The Extrasolar Planets Encyclopaedia, http://exoplanet.eu}. The majority of them are very massive and on very short period orbits, something that was quite unexpected from the theories of giant-planet formation. However, since a few years, giant planets much more similar to the solar system giants \citep[e.g.][]{Wright-2008} as well as super-Earth planets have been detected (mass from 2 to 10 M$_{\oplus}$) \citep[e.g.][]{Mayor-2009a, Mayor-2009b, Udry-2007b}. This has become possible thanks to three important improvements of the RV technique. First, the precision of spectrographs improved considerably, reaching now a 100\,cm\,s$^{-1}$ precision level \citep[typically 100 cm\,s$^{-1}$ in 1 minute for a $V=7.5$ K0 dwarf, using the HARPS spectrograph on the ESO 3.6 meter telescope, see][]{Pepe-2005}. A second improvement has been achieved through an appropriate observational strategy, which allows us to average out perturbations caused by stellar oscillations \citep[][]{Santos-2004a}. Finally, several years of follow-up helped us find long-period planets as well as very small mass ones. At the 100\,cm\,s$^{-1}$ level of accuracy, we start to be confronted with noises caused by the stars themselves. Stellar noise is the result of three types of perturbation produced by three different physical phenomena: oscillations, granulation, and magnetic activity. 

Oscillations of solar type stars, which can be seen as a dilatation and contraction of external envelopes over timescales of a few minutes \citep[5 minutes for the Sun;][]{Schrijver-2000,Broomhall-2009}, is caused by pressure waves (p-modes) that propagate at the surface of solar type stars. The individual amplitudes of p-modes are typically from a few to tens of cm\,s$^{-1}$, but the interference of tens of modes with close frequencies introduce RV variations of several cm\,s$^{-1}$, depending on the star's spectral type and evolutionary stage \citep[][]{Bedding-2007a,Bouchy-2003,Bedding-2003,Schrijver-2000}. The amplitude and period of the oscillation modes increase with mass along the main sequence. Theory and observations show that the frequencies of the p-modes rise with the square root of the mean density of the star and that their amplitudes are proportional to the ratio of the luminosity over the mass \citep[][]{Christensen-Dalsgaard-2004,OToole-2008}. 

The convection in external layers of solar type stars drives different phenomena of granulation (granulation, mesogranulation, and supergranulation), which also affect the RV measurements on time scales going from several minutes to several hours. In order of timescale and size of convective pattern, we first have granulation, whose typical timescale is shorter than 25 minutes \citep[][]{Title-1989,Del_Moro-2004b}. Then comes mesogranulation and finally supergranulation, with timescales of up to 33 hours \citep[][]{Del_Moro-2004a}. When integrated over the entire stellar disk, all these convection perturbations present a noise level on the order of the meter-per-second.

On longer timescales, similar to the star rotational period, the presence of activity related spots and plages perturbs precise RV measurements. Spots and plages on the surface of a star will break the flux balance between the red-shifted and the blue-shifted halves of the star. As the star rotates, a spot group, or a plage,  moves across the stellar disk and produces an apparent Doppler shift \citep[][]{Saar-1997,Queloz-2001,Huelamo-2008,Lagrange-2010}. This effect can be hard to distinguish from the signal caused by the presence of a planet. For the Sun at maximum activity of cycle 23, \citet[][]{Meunier-2010a} find a noise related to spot groups and plages of 42\,cm\,s$^{-1}$. Because the temperature of spot groups and plages are different from the mean stellar surface, and because active regions contain both, the noise induced by them will usually be compensated, but not entirely, because the surface ratio between spots and plages varies \citep[e.g.][]{Chapman-2001}. According to \citet{Meunier-2010a}, the major perturbative effect of activity on RVs is not the one induced directly by spot groups and plages, but the one caused by the inhibition of convection in active regions \citep[e.g.][]{Dravins-1982,Livingston-1982,Brandt-1990,Gray-1992}. This effect leads to a blueshift distortion of spectrum lines, which results in a noise varying from 40\,cm\,s$^{-1}$ at minimum activity to 140\,cm\,s$^{-1}$ at maximum.

The Earth produces a radial-velocity perturbation of 9\,cm\,s$^{-1}$ on the Sun, which would be completely masked by the stellar noise. If we manage to understand the structure of these different kinds of noise, we could use appropriate observational strategies to reduce the stellar noise contribution as much as possible, and thus find very small mass planets far from their host star. This investigation of optimized observational strategies is essential for future and more accurate instruments, such as ESPRESSO@VLT (http://espresso.astro.up.pt/, precision expected: 10\,cm\,s$^{-1}$) or CODEX@E-ELT \citep[e.g.][precision expected: 2\,cm\,s$^{-1}$]{Pasquini-2008}, in order to detect Earth twins (Earth-mass planets in habitable regions).

The two first types of noise, which are caused by oscillations and granulation phenomena, have been discussed in a previous paper \citep[][hereafter Paper I]{Dumusque-2010a}. Starting from HARPS asteroseismology measurements, we characterized these two kinds of noise using the power spectrum representation. Then we derived an optimized observational strategy, reducing at best the stellar noise contribution while keeping a reasonable total observational time.

In the present paper, we continue this first study and include the noise coming from magnetic activity spot groups. Starting from observations of the Sun, which is the only star where we can resolve spot groups, we simulate the appearance of spot groups on its surface and calculate the RV contribution. Adding the noise coming from magnetic activity to the results obtained in Paper I, we develop an observational strategy that simultaneously reduces the three kinds of stellar noise. To finish, we calculate the corresponding detection limits, as well as the expected number of planets that could be found by comparing our results with Bern's model of planetary formation.

\section{Simulation of different activity levels}\label{sec:1}

For decades surveys have studied the activity level of FGK stars in the solar neighborhood \citep[e.g. ][]{Baliunas-1995, Hall-2007}. 
%The level of activity is most often derived using the Ca II H and K lines, which vary with the chromospheric emission.
One of the main results of these surveys is the bi-modality of the activity level distribution \citep[e.g. ][]{Vaughan-1980, Henry-1996}. About 70\% of the stars, even those with a clear Sun-like activity cycle, seem to have an activity level below $\log(R'_{HK})=-4.75$ \citep[see][for more information about the chromospheric emission ratio, $R'_{HK}$]{Noyes-1984}. The others have a higher activity level, and no crossing between the \emph{low-activity} and \emph{high-activity} region seems to occur. Early results of the Kepler mission arrive at the conclusion that there may be more active stars than expected \citep[see][]{Basri-2010a}. Of more than 100000 stars with a surface gravity, $\log g$, higher than 4 and an effective temperature between 3200 and 19000 K, 50\,\% seem to be more active than the active Sun. Nevertheless, very low-activity stars have been discovered in this sample. 

In the present paper, we simulate the activity of stars below $\log(R'_{HK})=-4.75$. The goal is to reproduce as realistically as possible the appearance of stellar spot groups on the surface of these stars. Because the Sun is the only star with an activity level between $-5$ and $-4.75$ for which we can resolve stellar spot groups, we will use it as a proxy for \emph{low-activity} stars. Stars with a $\log(R'_{HK})$ higher than -4.75 will not be simulated. First of all because of a lack of information about the activity phenomenon of these stars, and secondly, because the activity noise will probably be too important to find very small mass planets in habitable regions with the RV technique.

\subsection{Activity of the Sun}\label{subsec:1.1}

The Sun has a 22-year magnetic cycle as well as a 11-year sunspot cycle. During these 11 years, the activity level varies between $\log(R'_{HK})=-5$ at minimum and $\log(R'_{HK})=-4.75$ at maximum. At minimum there is no spot group on the surface of the Sun. When the activity level rises, several spot groups start to appear at a latitude of about $\pm30^{\circ}$. Before reaching the maximum activity level, the number of spot groups will increase progressively, whereas the latitude of spot groups will decrease ($\pm15^{\circ}$ at maximum activity). After this maximum phase, the migration of spot groups toward the equator continues and the number of groups decreases progressively until they all disappear. At the end of a cycle, the activity level returns to $\log(R'_{HK})=-5$, and another sunspot cycle of 11-year begins again. 

Because the latitude of the spot groups changes during the sunspot cycle, it is important to recall that the Sun has a differential rotation in latitude. Therefore, spot groups rotate faster at the equator than at the poles, according to the equation
\begin{equation}\label{eq:1}
\omega=A+B\sin^2\theta,
\end{equation}
where $\omega$ is the angular speed in $degree / day$, $\theta$ the latitude, and A and B are two constants equal to $14.476 \pm 0.006$ and $-2.875 \pm 0.058$ \citep[see][]{Howard-1996}.

During the 11-year sunspot cycle or an even longer time, the spot groups do not appear randomly in longitude, but 15$^{deg}$ around preferred \emph{active longitudes} \citep[e.g.][]{Ivanov-2007, Berdyugina-2003}. These preferred longitudes turn with the Sun's differential rotation and are observed for large spot groups\footnote{Large spot groups represent 11\% of the total number of spot groups observed, corresponding to 67\% of the total spot group area.}. Observation of the Sun over several sunspot cycle has shown that
\begin{itemize}
\item there are  normally two \emph{active longitudes} in the northern hemisphere and two in the southern one,
\item the two \emph{active longitudes} are shifted by 180$^{deg}$ in each hemisphere,
\item the \emph{active longitudes} in the south are shifted by 90$^{deg}$ in regard to those in the north,
\item sometimes, each hemisphere can have a third \emph{active longitude}, which is shifted by 90$^{deg}$ with regard to the two normal ones.
\end{itemize}

When sunspots appear, their diameter is very small, approximately 0.003$\,R_{\odot}$. The majority of them will disappear in a couple of hours, but a few will evolve and become bigger spots, which will gather together as spot groups. These can last for several days to a few months. According to \citet{Howard-2000}, the lifetime distribution of spot groups is
\begin{itemize}
\item 50\% of the spot groups have a lifetime between 1 hour and 2 days,
\item 40\% of the spot groups have a lifetime between 2 days and 11 days,
\item 10\% of the spot groups have a lifetime between 11 days and 2 months.
\item For the Sun at maximum activity ($\log(R'_{HK})=-4.75$), there is one group spot per sunspot cycle that can last from 2 to 6 months.
\end{itemize}
During its lifetime, a spot group will rapidly increase in area until the maximum size is reached, after which the decrease will follow in a slower way \citep[e.g][]{Howard-2000}. The increase and decrease times are not precisely defined. For our simulation, we will set these times to 1/3 and 2/3 of the total lifetime of a spot group.

The maximum spot group filling factor\footnote{The filling factor is just the ratio of the area of the spot group over the area of the visible hemisphere.}, $f_{s,max}$, is linked to the lifetime of the spot group, $T$, by the equation \citep[][]{Howard-2000}
\begin{equation}\label{eq:2}
f_{s,max}=10^{-5}T,
\end{equation}
where $f_{s,max}$ is expressed in hemisphere and $T$ in days. 

Once the lifetime and the maximal size of spot groups are defined, we need a probabilistic process to mimic the observed appearance of spot groups on the solar surface. According to \citet{Hoyt-1998}, the appearance of spot groups can be described by a Poisson law. We can therefore define the probability, $P$, to have a certain number of spot groups, $N$, by the formula
%With the hypothesis that the appearance of spot groups is a phenomenon independent of time and that we know their average of appearance, we can use a Poisson law to describe the evolution of spot groups. On a short periods according to the 11-years sunspot cycle, the hypothesis above are respected., therefore, the appearance of spot groups can be describe by the following Poisson process:
\begin{equation}\label{eq:3}
P[(N(t+\tau)-N(t))=k]=\frac{e^{-\lambda\tau}(\lambda\tau)^k}{k!} \qquad k=0, 1, \cdots,
\end{equation}
where $t$ is the time, $\tau$ the time step and $\lambda$ the average appearance of spot groups per unit of time.

\begin{figure}
\begin{center}
\resizebox{\hsize}{!}{\includegraphics{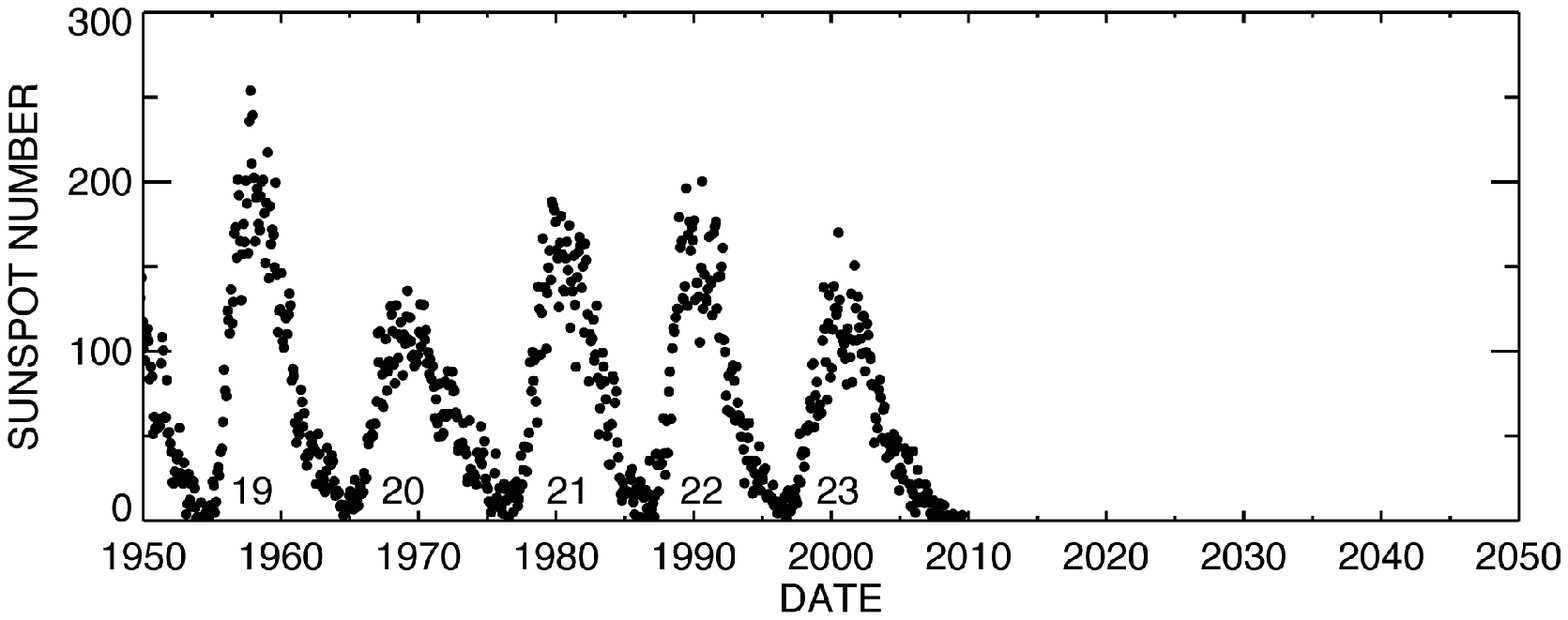}}
\resizebox{\hsize}{!}{\includegraphics{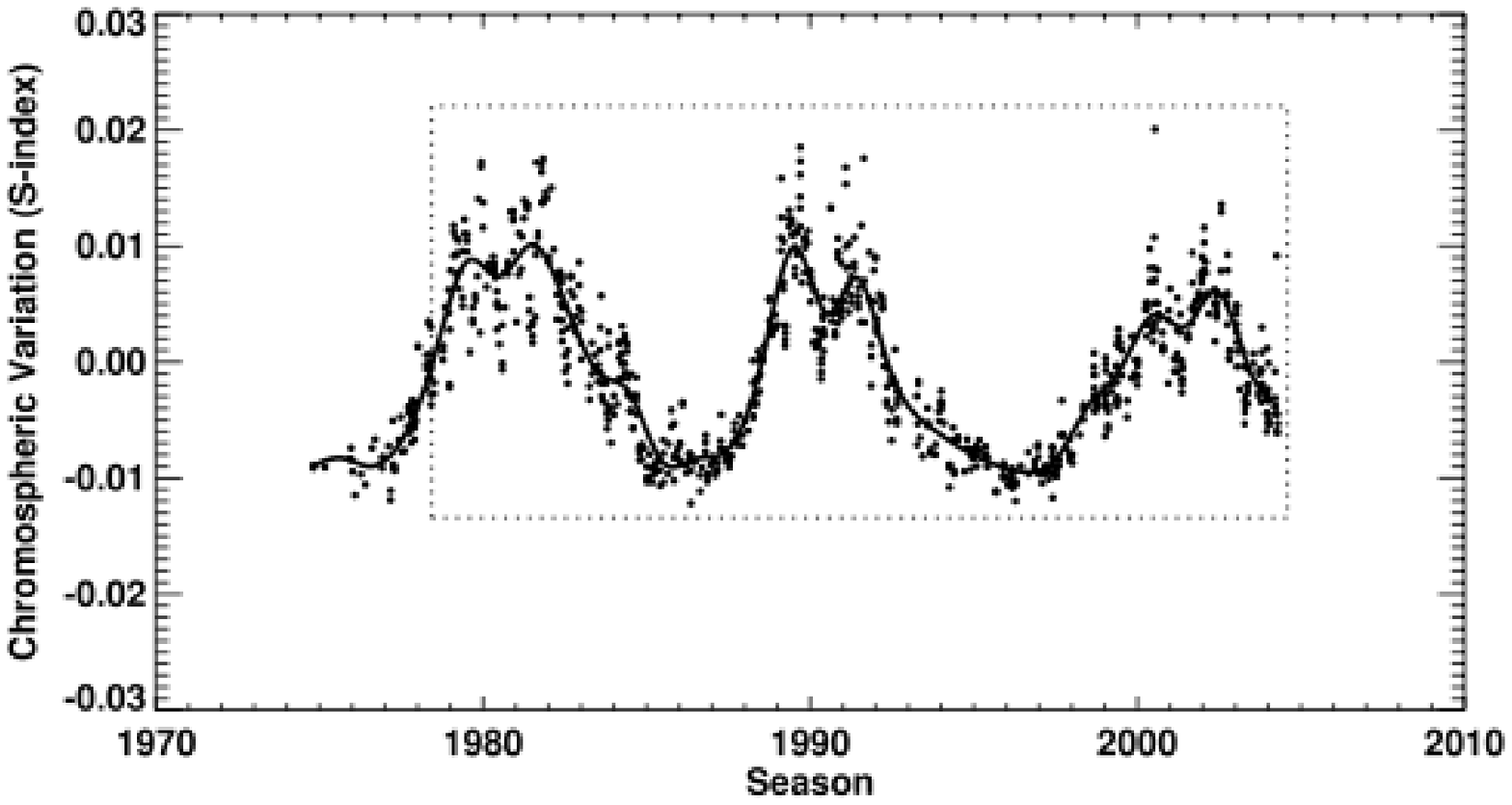}}
\caption{\emph{Upper Panel: }Monthly averages of the sunspot numbers. We clearly see the 11-year sunspot cycle (from http://solarscience.msfc.nasa.gov/SunspotCycle.shtml). \emph{Lower Panel: } Variation of the S-index as a function of time for the Sun \citep[from][]{Lockwood-2007}.}
\label{fig:1}
\end{center}
\end{figure}
The upper panel in Fig. \ref{fig:1} represents the continuous daily observation of sunspots (Marshall Space Flight Centre, http://solarscience.msfc.nasa.gov/SunspotCycle.shtml). Clearly, the number of spots varies with the 11-year sunspot cycle of the Sun. Comparing the two graphs in Fig. \ref{fig:1}, we notice that the number of spots is correlated with the S-index and thus with the $\log(R'_{HK})$ (because S and $\log(R'_{HK})$ are proportional, see \citet{Noyes-1984}). In conclusion, no spot corresponds to $\log(R'_{HK})=-5$, and the maximum number of spots corresponds to $\log(R'_{HK})=-4.75$. According to this argument, we decide to simulate three different levels of activity:
\begin{itemize}
\item $\log(R'_{HK})=-5$: the Sun is at minimum activity and there is no spot group on the solar surface. Therefore, no noise from the spot groups perturbs the RVs (case of Paper I).
\item $\log(R'_{HK})=-4.75$: the Sun is at maximum activity. The number of spots in the visible hemisphere is approximately equal to 150 and they are located at $\pm15^{\circ}$ of latitude \citep[e.g][]{Howard-2000}.
\item $\log(R'_{HK})=-4.9$: the Sun is at medium activity, which corresponds to a total number of 75 spots at $\pm22.5^{\circ}$ of latitude in the visible hemisphere.
\end{itemize}

\begin{figure}
\begin{center}
\includegraphics[width=8cm]{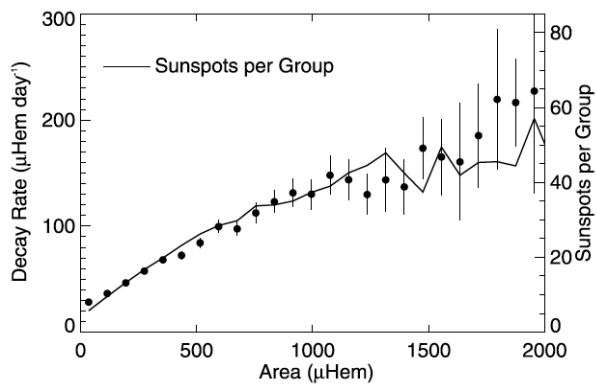}
\caption{Number of sunspots per spot group (right scale) as a function of the spot group maximal area in millionth of hemisphere 
%To have the maximal area in filling factor, we have to divide by $10^{-6}$. The solid line shows the average number of sunspots per group as a function of area on the scale given on the right 
\citep[source:][]{Hathaway-2008}.}
\label{fig:2}
\end{center}
\end{figure}

The activity level is given by a defined number of spots, but the properties defined above, such as the lifetime distribution, the maximum filling factor (Eq. \ref{eq:2}), and the appearance statistical law (Eq. \ref{eq:3}), are given for spot groups. In order to use all these properties, we need to know the number of spot groups of a defined number of spots. This can be done using the work of \citet{Hathaway-2008}, which gives the number of sunspots per spot group as a function of the spot group maximum filling factor (see Fig. \ref{fig:2}). 

\subsection{Activity simulation}\label{subsec:1.2}

In order to check our ability to detect Earth-like planets through activity effects, our simulations will try to mimic the Sun activity phenomena as much as possible.

For the simulation, we start with the following initial conditions:
\begin{itemize}
\item We set the number of spots, $N_s$, for a given activity level. 
\item Then, according to the lifetime distribution described in Sect. \ref{subsec:1.1}, we create a first spot group and calculate its maximum filling factor using its lifetime (Eq. \ref{eq:2}). \item This maximum filling factor gives us, using Fig. \ref{fig:2}, the number of spots present in this spot group, $N_{s,\,group}$. If $N_{s,\,group} < N_s$, we create a second spot group and so on, until $\sum{N_{s,\,group}} = N_s$. 
\item Each spot group is put on the latitude corresponding to the activity level and on the different \emph{active longitudes}, with no possible overlap. The longitudes of the spot groups are selected using a Gaussian distribution\footnote{Because 68\,\% of the values generated by a Gaussian distribution will fall in the range $\mu \pm \sigma$, where $\mu$ is the mean and $\sigma$ the dispersion, this technique allows us to put 68\% of the total spot groups area at $15^\circ$ around the \emph{active longitudes} (67 \% observed, see Sect. \ref{subsec:1.1}).} centered on \emph{active longitudes} and with a dispersion of $15^{deg}$.
\end{itemize}

Once the initial conditions are created, we start the time evolution of the system, using a rotational period of the Sun at the equator of 26 days. The time step of the simulation is set to two hours. All spot groups begin with a null area and increase their size linearly until they reach the maximum after 1/3 of their lifetime (see Eq. \ref{eq:2}). The decreasing phase is linearly during the remaining 2/3 of their lifetime. The number of new spot groups at each time step is given by the appearance statistical law (see Eq. \ref{eq:3}). 

At the beginning of the simulation, the value for the average appearance of spot groups, $\lambda$, is unknown. To find the right value, we begin the simulation with an arbitrary number and calculate the total filling factor for each time step. Using Fig. \ref{fig:2}, this filling factor give us the total number of spots. If the mean total number of spots in not equal to $N_s$ in several years, we restart the simulation with another value for $\lambda$. Finally, after several tests, $\lambda$ is set to 2.4 and 4.8 for an activity level of $\log(R'_{HK})=-4.9$ and $\log(R'_{HK})=-4.75$, respectively.

For each level of activity, we generate 100 simulations with an equal probability of having two or three \emph{active longitudes} per hemisphere. 
%The output files gives us, for each time step and each spot group, the position and size of the spot group, as well as the number of spots it contains.

\subsection{Radial-velocities induced by activity-related spot groups}\label{subsec:1.3}

\begin{figure}
\begin{center}
\includegraphics[width=9.5cm]{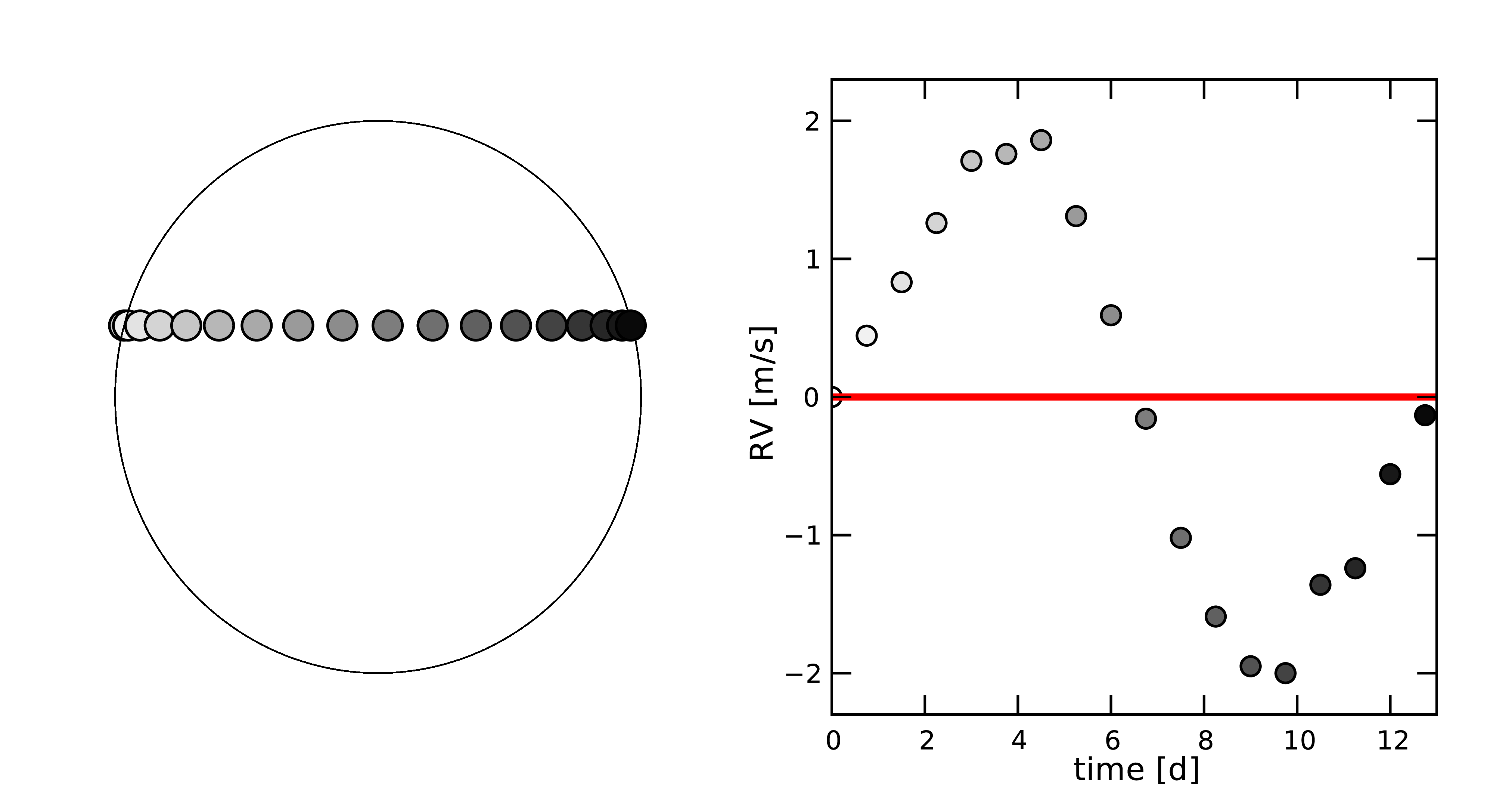}
\caption{RVs produced by a spot group at $15^{deg}$ of latitude, with a filling factor of 0.001. The simulation is made using the program SOAP.}
\label{fig:3}
\end{center}
\end{figure}

When spot groups are present on a rotating star, they break the flux balance between the red-shifted and the blue-shifted halves of the star. As the star rotates, a spot group moves across the stellar disk and produces apparent Doppler shifts. This modulation is often hard (sometimes impossible) to distinguish from the Doppler modulation caused by the gravitational pull of a planet. To model the radial-velocity variations imposed by stellar spot groups, we used the program SOAP (Bonfils \& Santos, in prep.). SOAP computes the rotational broadening by sampling the stellar disk on a grid. Each grid cell is assigned a Gaussian function that represents the typical emerging spectral line (or, equivalently, the spectrum's Cross Correlated Function (CCF)). All cells are Doppler-shifted according to their projected velocities toward the observer's line-of-sight, and averaged with a weight following a linear limb-darkening law ($\alpha=0.6$). When spot groups are added to the stellar surface, SOAP computes which grid cells they occupy and change the weight for those given cells. The weight can be set to zero for a dark spot group, to a fraction of the stellar brightness if we assume a given brilliance for the spot group, or to more than the stellar brightness to simulate plages. Finally, for a given spot group configuration, SOAP delivers the modified spectral line, the flux, the radial-velocity, and the bisector span as a function of the star's rotating phase. Note that SOAP merely calculates the flux effect induced by spot groups. It does not model any type of granulation, therefore, the inhibition of convection (granulation) in active regions such as spot groups is not took into account \citep[][]{Meunier-2010a}.
\begin{figure}
\begin{center}
\resizebox{\hsize}{!}{\includegraphics{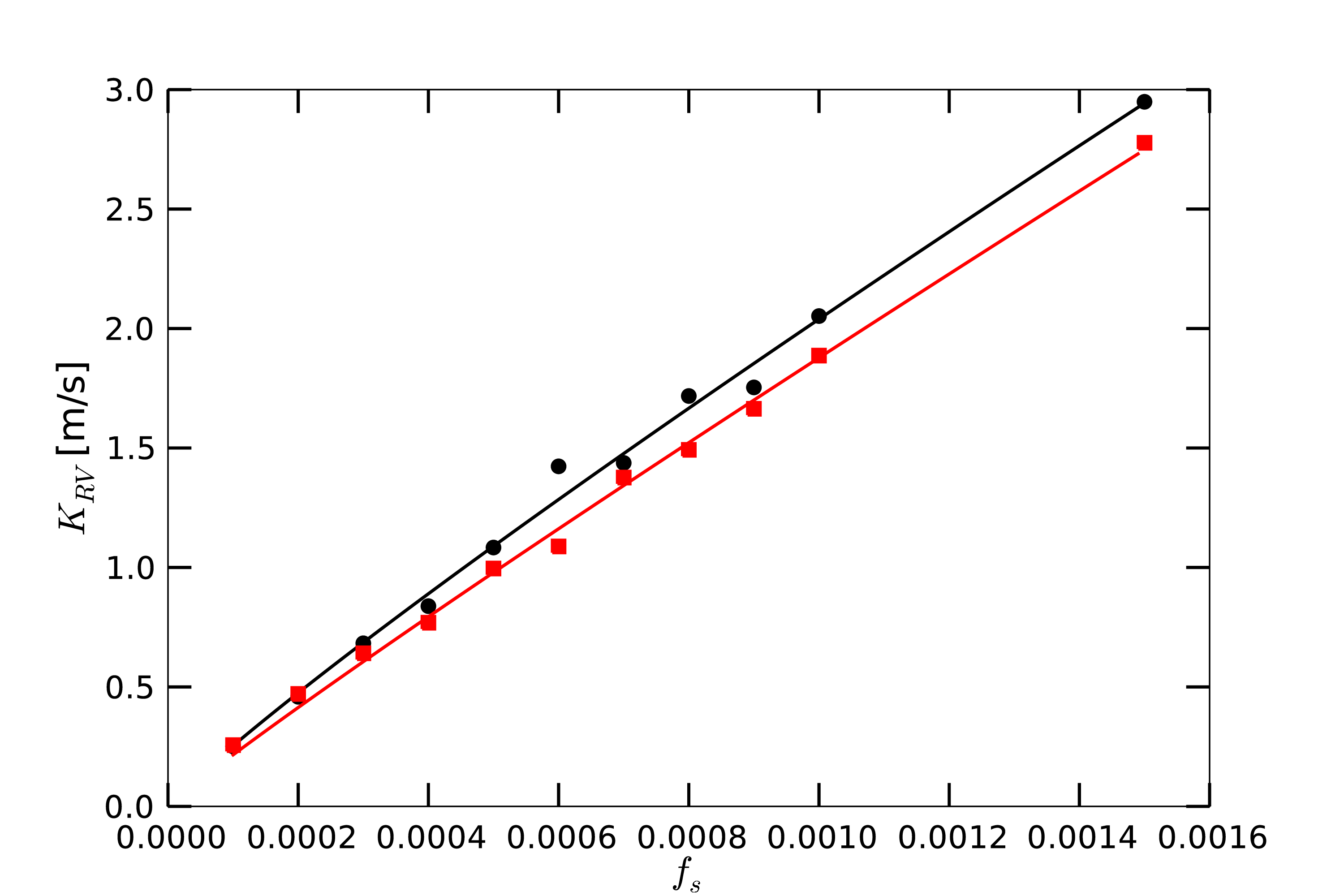}}
\caption{Semi-amplitude of the RV signal, $K_{RV}$ induced by a spot group as a function of the spot group filling factor. These results have been obtained with the program SOAP. We can see in black the result for spot groups at a latitude of 15$^{deg}$, and in gray (red), of 22.5$^{deg}$.}
\label{fig:3-1}
\end{center}
\end{figure}
\begin{figure*}[!ht]
\begin{center}
\includegraphics[width=8cm]{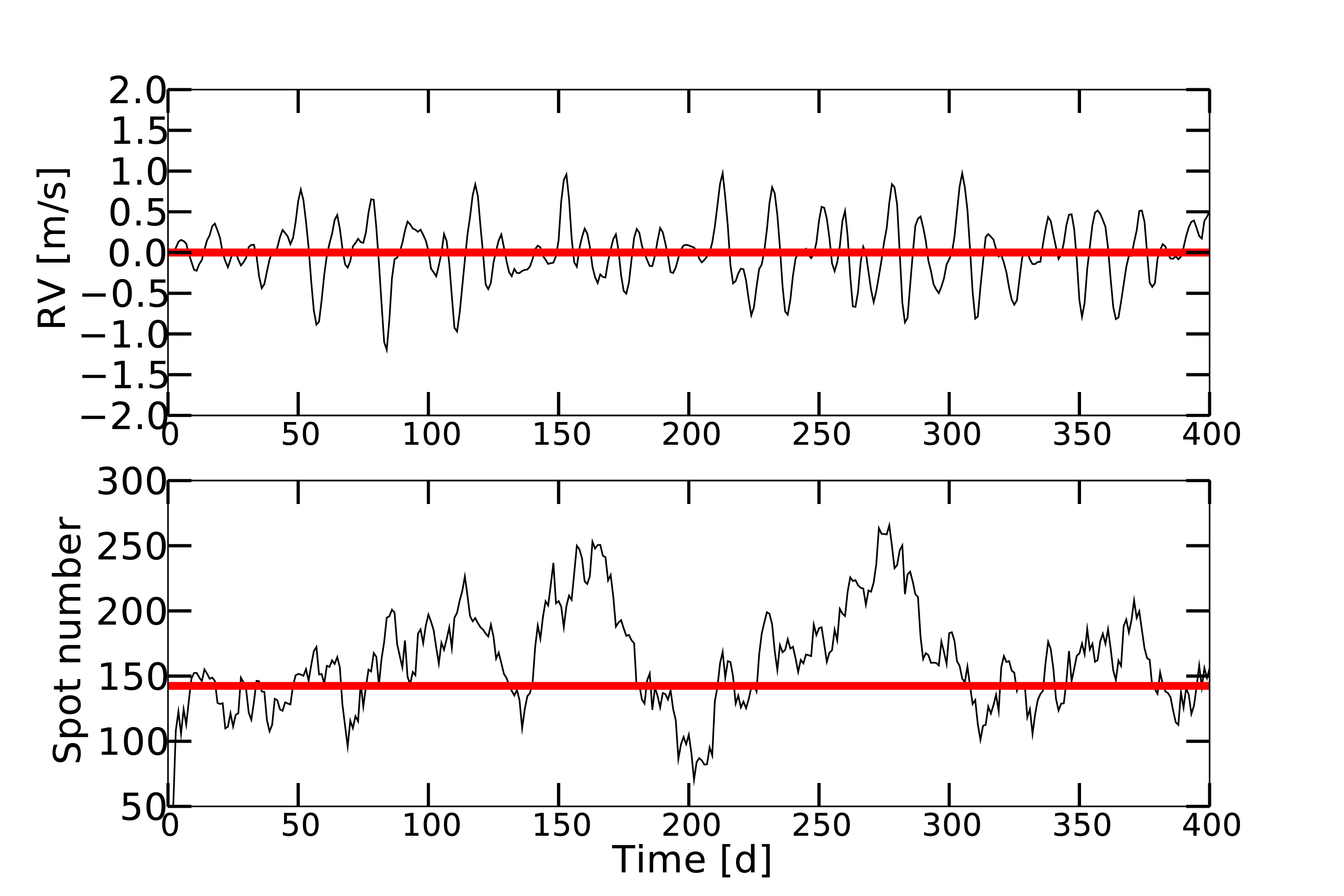}
\includegraphics[width=8cm]{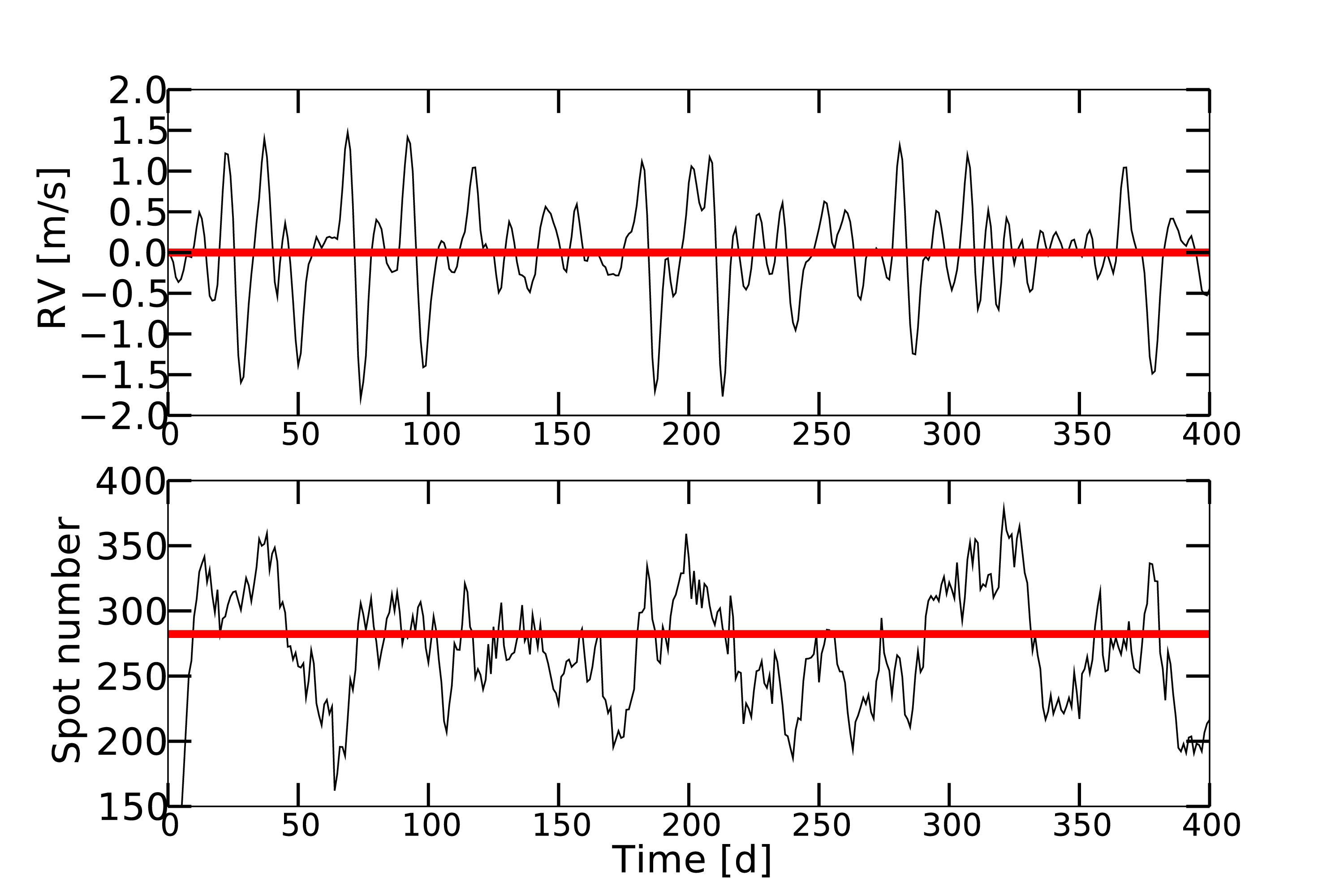}
\caption{Simulations for an equatorial rotational period of 26 days and an activity level $\log(R'_{HK})=-4.9$ (left panel) and for $\log(R'_{HK})=-4.75$ (right panel). The horizontal line on the top and bottom graphs are the RV mean and the spot number mean over the 4 years of the simulation, respectively. Just the 400 first days are shown here for clearness.}
\label{fig:4}
\end{center}
\end{figure*}

Normally a spot group is constituted of several small spots. For the simulation we considered each spot group as a unique big spot to facilitates the operation. We used the program SOAP to calculate the RV effect induced by a spot group presenting a filling factor, $f_{s1}$ of 0.001 and a latitude of 15$^{deg}$ (Fig. \ref{fig:3}) or 22.5$^{deg}$. The first latitude corresponds to a $\log(R'_{HK})$ of -4.75 and the second to a $\log(R'_{HK})$ of -4.9. To get the RV effect for any spot group filling factor, $f_{s2}$, we first calculate the RV semi-amplitude with SOAP, $K_{RV}$, which is induced by different spot group sizes. Then we fitted a power law model to these data (Fig. \ref{fig:3-1}) and found that
\begin{itemize}
\item for a spot group at a latitude of 15$^{deg}$, $K_{RV} \propto f_s^{0.90}$,
\item for a spot group at a latitude of 22.5$^{deg}$, $K_{RV} \propto f_s^{0.94}$.
\end{itemize}
These results are consistent with the relation presented in \citet{Saar-1997}, where the authors have found that $K_{RV} \propto f_s^{0.90}$. We will use this relation in our simulation for all latitudes. Supposing that besides the amplitude, the RV signal has the same shape for any spot group size, we calculate the proportionality coefficient between two signals, which is $K_{RV2}/K_{RV1}=(f_{s2}/f_{s1})^{0.9}$. Multiplying the RV effect induced by a spot group with $f_s=0.001$ by this coefficient yields the RV effect induced by any spot group size.

The simulation does not consider spot groups with a lifetime shorter than one day, because the maximum filling factors will be very small an therefore the RV contribution negligible.

We present here a simulation for the Sun, which is seen equator on, therefore the inclination angle between the line of sight and the rotation axis of the star, $i$, equals 90$^{\circ}$. For other stars, this assumption cannot be made because for the majority of them we do not have access to the $i$ angle. Moreover, when $i$ decreases, the RV which is equal to $v\sin{i}$ decreases as well . Thus the RV effect of spot groups crossing the visible stellar disc, which is proportional to $v\sin{i}$, will be lower. In conclusion, considering $i=90^{\circ}$ in the simulation gives us the maximum RV effect that spot groups can induce.

\subsection{Comparison with other works}\label{subsec:1.4}

\begin{figure}
\begin{center}
\includegraphics[width=8cm]{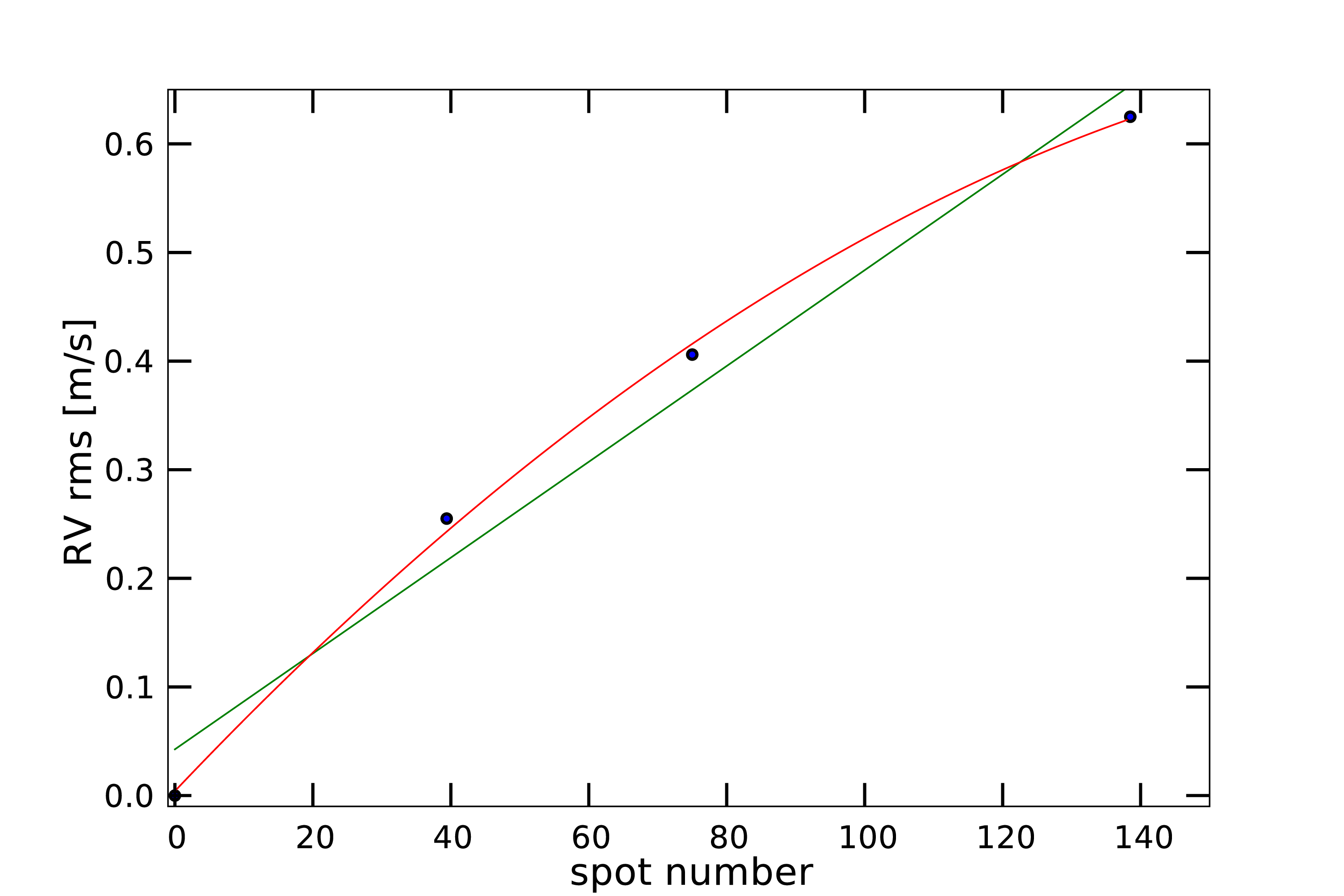}
\caption{Evolution of the RV variation as a function of the total number of spots. Each point corresponds to 100 simulations.}
\label{fig:4.1}
\end{center}
\end{figure}

An example of simulations for $\log(R'_{HK})$=-4.9 and -4.75 can be seen in Fig. \ref{fig:4}. The amplitudes found can be compared to a recent result from \citet{Meunier-2010a}, who calculated the RV induced by the real distribution of spot groups during the solar sunspot cycle 23. For maximum activity, they find a RV variation of 48\,cm\,s$^{-1}$. Note that this variation is valid only for spot groups and was derived without taking into account plages or inhibition of the convection in active regions. 
%At our knowledge, it is the first paper that present such a result, at least at this level of precision. 
In our case, after 100 simulations of a high-activity level, we arrive at a value of 63\,cm\,s$^{-1}$. Our estimate is 30\,\% higher that the one found by \citet{Meunier-2010a}, which is based on real position and size of spot groups. This can be explained because \citet{Meunier-2010a} used observations of the solar cycle 23. This was a fairly quiet cycle, presenting at maximum no more than 100 sunspots in the visible hemisphere (see Fig. \ref{fig:1}). In our case, the maximum number of sunspots was based on cycles 21 and 22, which were more intense (150 sunspots in the visible hemisphere at maximum activity level). In Fig. \ref{fig:4.1} we show the calculated RV variation obtained for different values of the total spot number. A quadratic fit to the data points indicates a RV variation of 51\,cm\,s$^{-1}$ for 100 spots, which is very close to the value cited in \citet{Meunier-2010a}. Thus, our simulation, starting from very simple characteristics of spot groups such as the lifetime distribution, the spot number according to the activity level, and the Poisson process of spot group appearance, manages very well to represent the RV effect induced by spot groups. Note that the inhibition of convection in active regions is not yet included in our simulation, and according to \citet{Meunier-2010a}, this effect could be dominant. Therefore, the RV effect of the activity is not yet fully simulated and could be underestimated. The very good agreement comparing the spot flux effect between real Sun observations and the simulation indicates that our simple model of spot groups is realistic. Including the inhibition of convection in the simulation is in progress and will be the topic of a forthcoming paper. Our activity simulation presents an advantage compared with the work done by \citet{Meunier-2010a} in the sense that we can adapt it to other stars than the Sun, if basic characteristics of the spot groups are known.

\begin{figure*}
\begin{center}
\includegraphics[width=6cm]{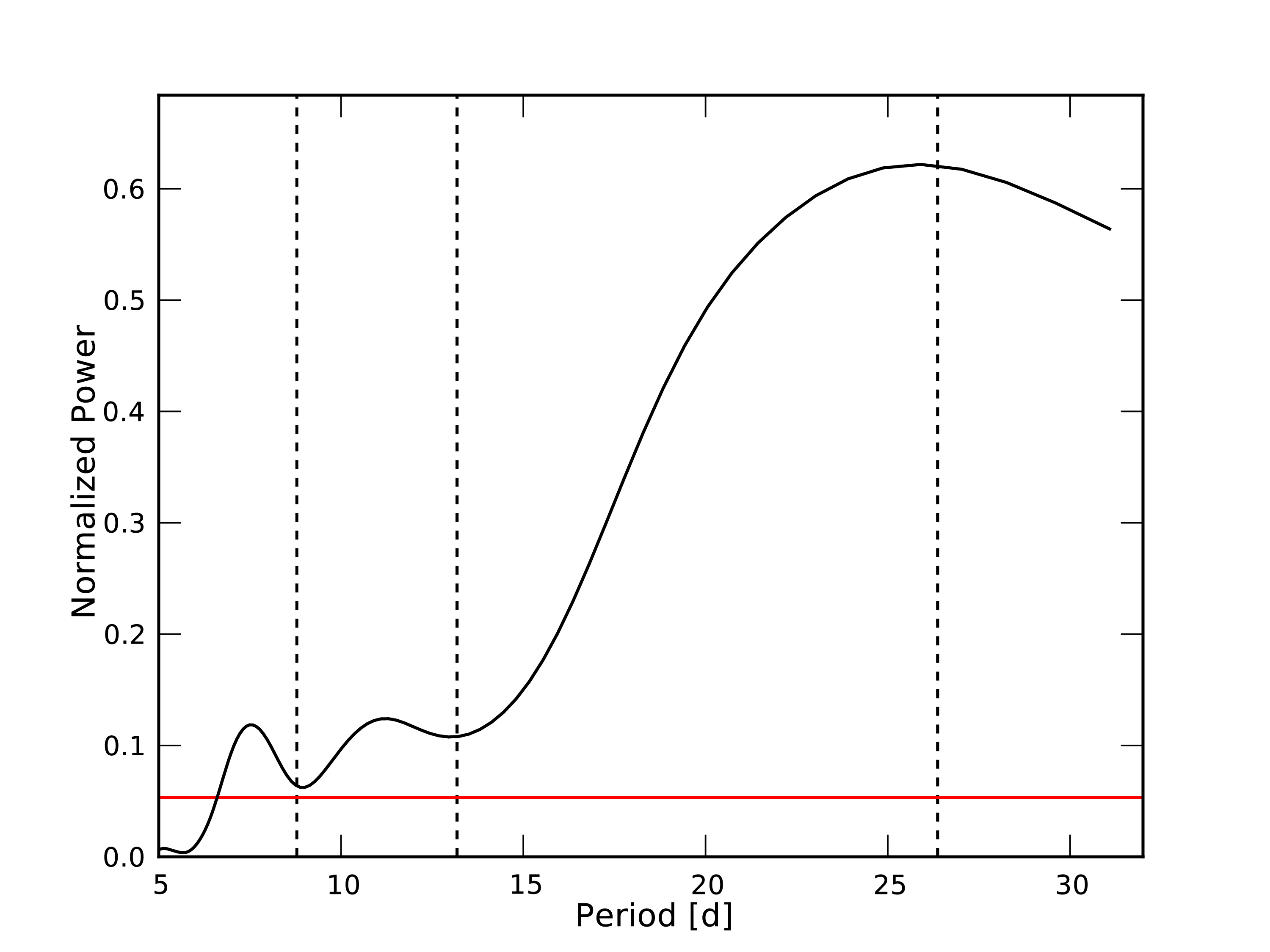}
\includegraphics[width=6cm]{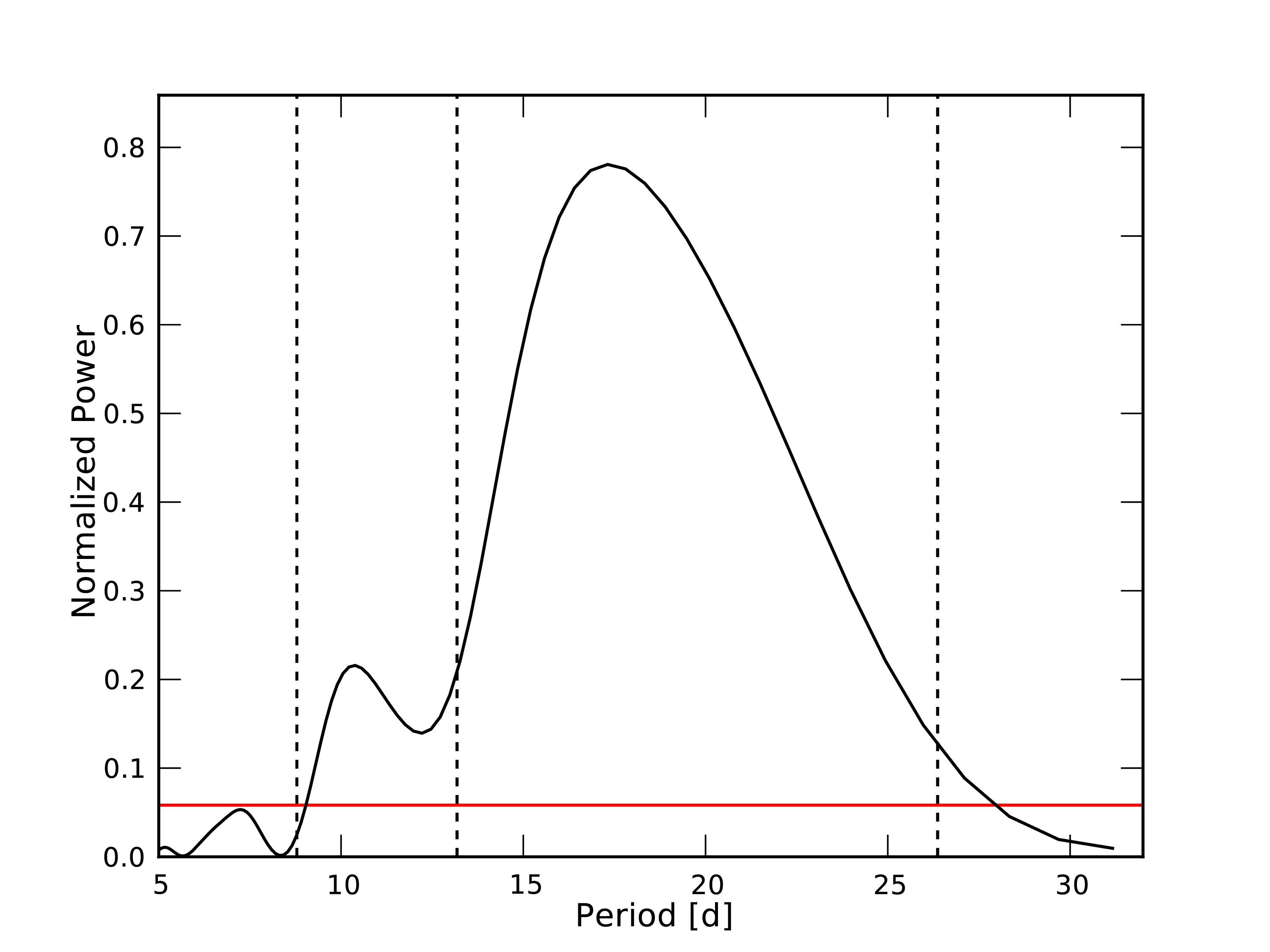}
\includegraphics[width=6cm]{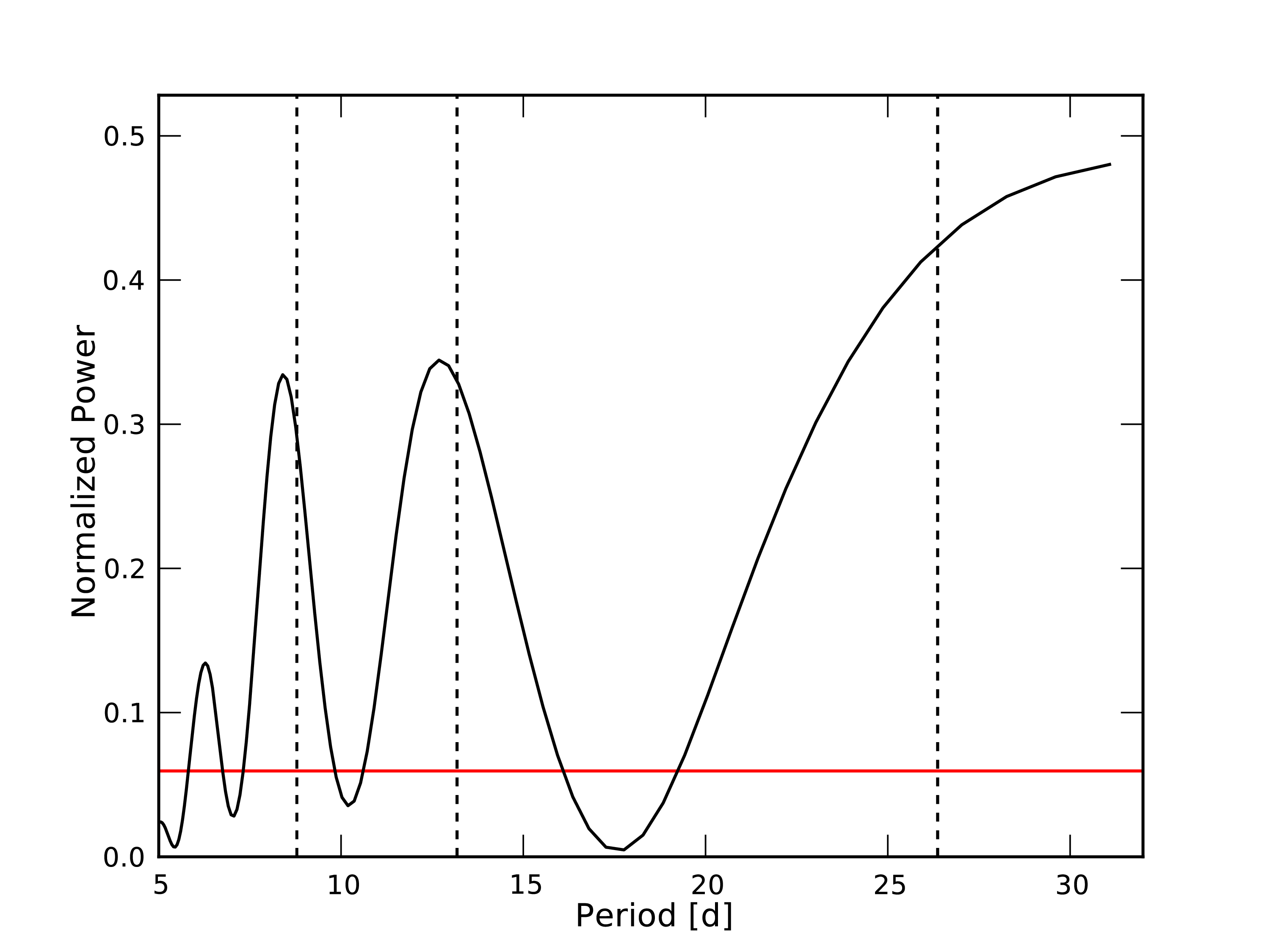}
\includegraphics[width=6cm]{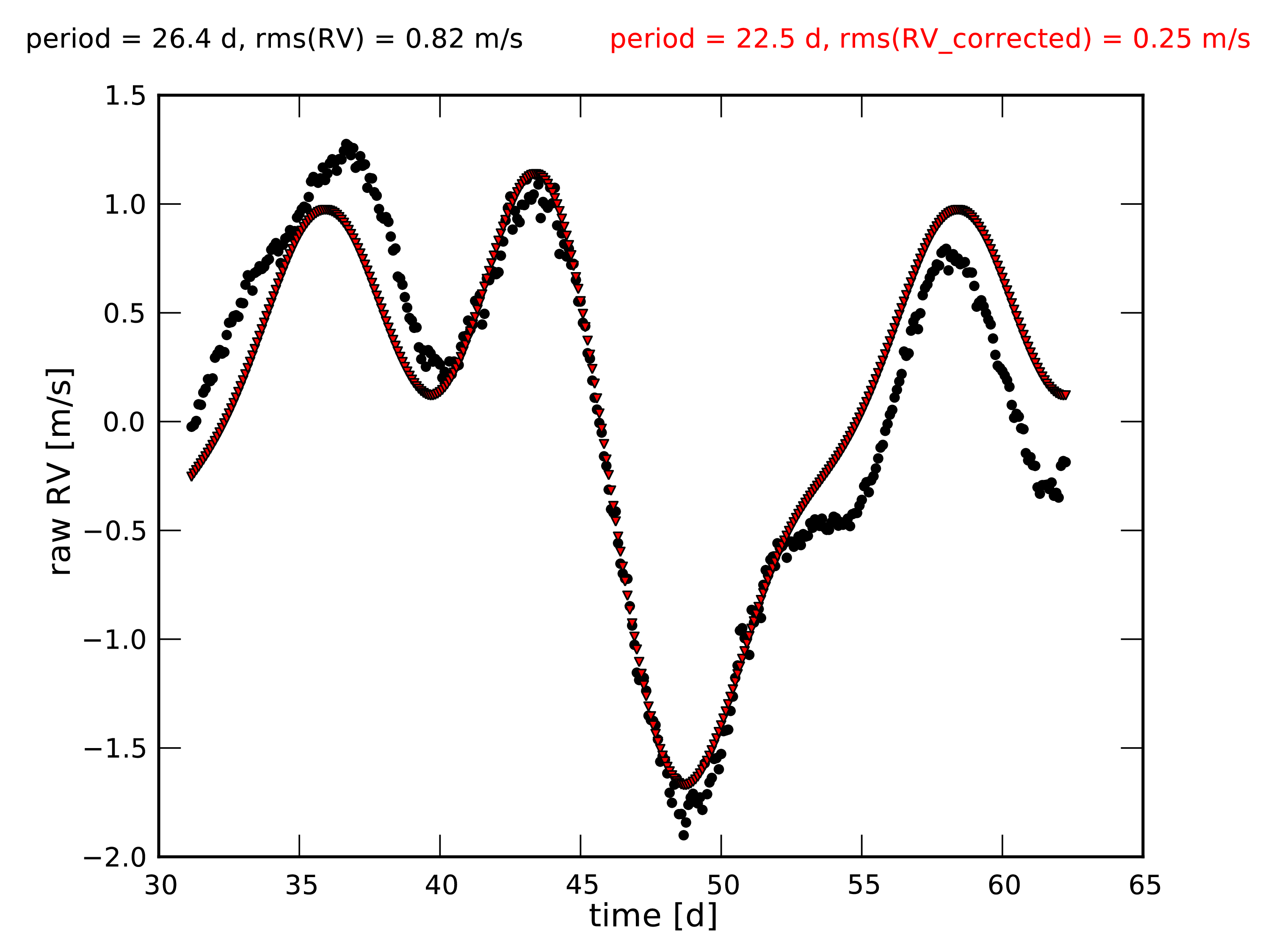}
\includegraphics[width=6cm]{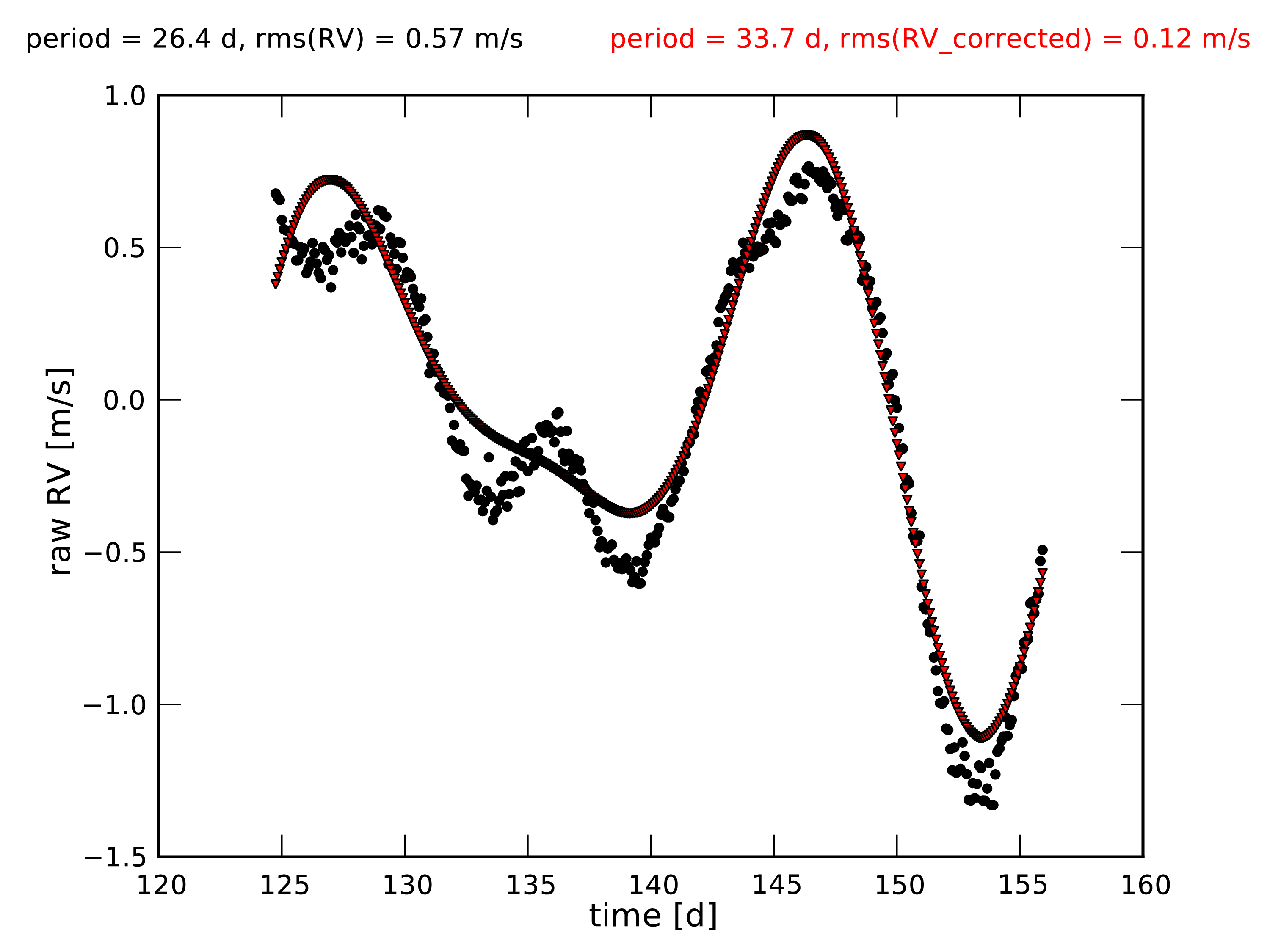}
\includegraphics[width=6cm]{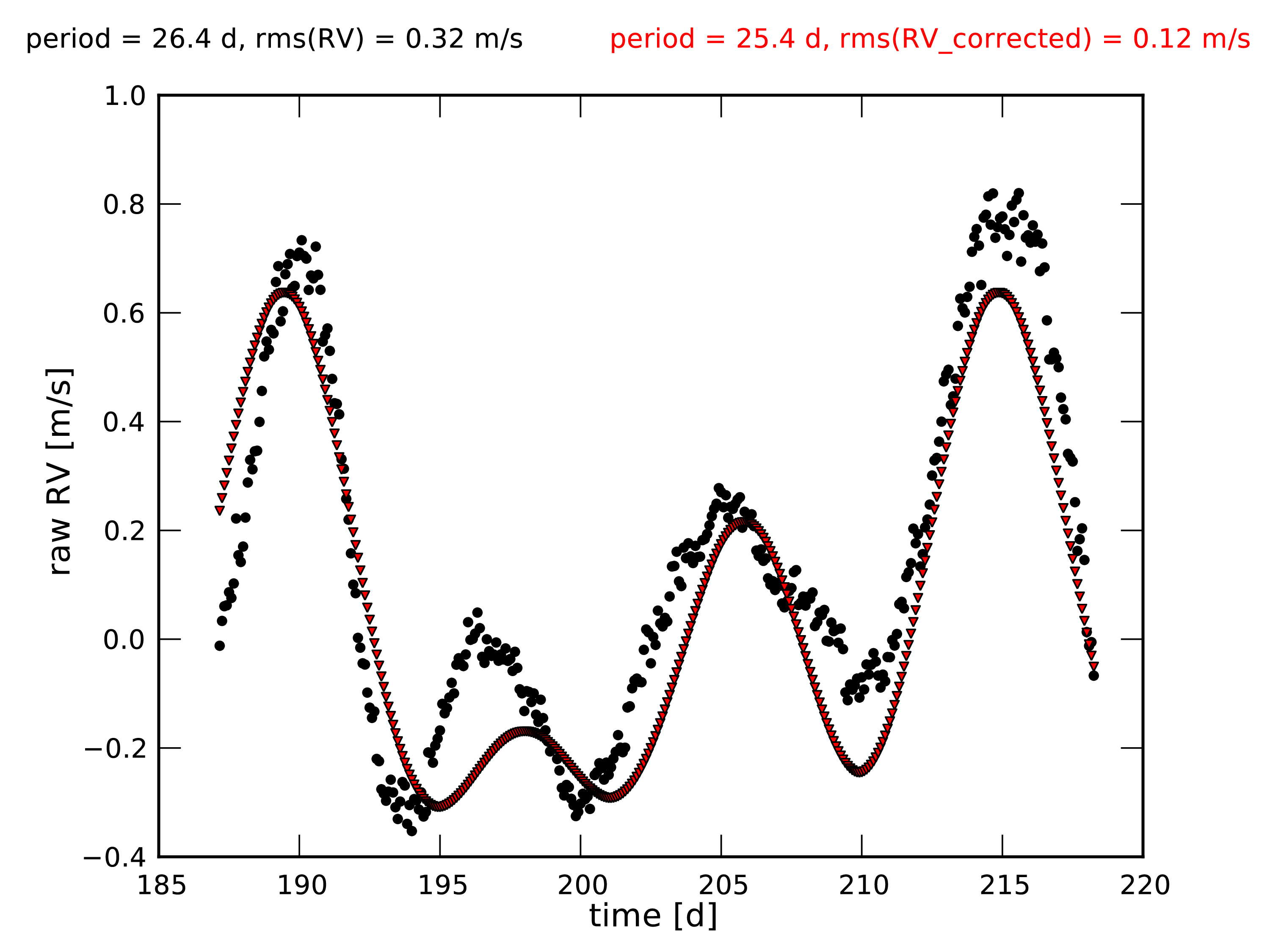}
\caption{Periodograms (\emph{top}) and corresponding RVs (\emph{bottom}) for three different slices of 31 days for the activity simulation shown in Fig. \ref{fig:4} ($\log(R'_{HK})=-4.75$). In the periodograms the horizontal line corresponds to the 1\%\,false-alarm probability and the vertical dashed lines to the rotational period of the star, $P_{rot}$, as well as $P_{rot}/2$ and $P_{rot}/3$. On the RV plots, the dots correspond to the simulated RVs and the triangles to a fitted function with three sin waves of period $P_{fit}$, $P_{fit}/2$ and $P_{fit}/3$ ($P_{fit}$ is a free parameter). $P_{rot}$ as well as the rms of the simulated RVs can be found in the upper left part of the plot. $P_{fit}$ as well as the rms of the corrected RVs (simulated minus fitted sine waves) can be found in the upper right part of the plot.}
\label{fig:4.0}
\end{center}
\end{figure*}

According to \citet{Queloz-2009} and \citet{Boisse-2010}, the RV signal caused by stellar spots is found at the rotational period of the star, $P_{rot}$, and its two first harmonics. When considering few big spots at the surface of the star that can change in size but not disappear rapidly, this assumption is valid because the activity signal stays in phase with rotation. In the two papers cited above, the authors consider active stars that are known to be dominated by few big spots that can live for several rotational periods. In our simulation, were we model non-active stars, we consider a lot of spots that are allowed to disappear rapidly and reappear at other places any time. This random behavior, ruled by a Poisson law, kills the activity signal phase even on timescales similar to the rotational period of the star. The RV signal however keeps a kind of periodicity on a rotational period timescale, but the period can be different form $P_{rot}$ and its two first harmonics. In Fig. \ref{fig:4.0} we plot the periodograms and the RVs for 31 day slices ($1.2\times P_{rot}$) of the simulation showed in Fig. \ref{fig:4} with an activity level of $\log(R'_{HK})=-4.75$. Looking at the periodograms, we see that sometimes important peaks are present at $P_{rot}$, $P_{rot}/2$ and $P_{rot}/3$ (plot on the right), as suspected by \citet{Queloz-2009} and \citet{Boisse-2010}. At other times, the peak at $P_{rot}$ is clear but nothing shows up at $P_{rot}/2$ and $P_{rot}/3$ (plot on the left) or at the rotational period and the two first harmonics (plot on the center). Looking now at the RV plots, we see that the model of three sine waves with period $P_{fit}$, $P_{fit}/2$ and $P_{fit}/3$ can fit the simulated RVs nicely. The comparison between the rms of the simulated RVs and the corrected one (simulated RVs minus fitted sine waves) shows that the three sine waves model reduces the activity noise by 70\,\%. Therefore this model can be used to correct short-term activity noise. However, if a small mass planet orbits the stars with a period close to $P_{rot}$, the three sine waves model will kill this signal too. We have therefore to be very careful when using these models, and only the study of other CCF parameters such as the full width at half maximum (FWHM) or the bissector span (BIS) will be able to give us clues on the nature of the signal: short-term activity or planet. This work will be done once the inhibition of the convective blueshift in active regions will be implemented into our model. In addition, the estimate of the rotational period using the three sine waves model, as shown in \citet{Queloz-2009}, is no longer valid. Indeed, we see that the fitted period, $P_{fit}$, can vary between 22 to 34 days, whereas the true rotational period of the star is fixed to 26.4 days in the simulation.

\section{Efficient observational strategies}\label{sec:2}

The RV signal induced by a spot group can be described as the succession of null contribution when the spot group is in the hidden hemisphere, and as sinusoidal shapes when the spot group moves across the visible hemisphere (Fig. \ref{fig:3}). This RV signal is regular in time and is composed of sinusoidal signals with periods equal to the rotational period of the star and its first harmonics \citep[][]{Boisse-2010}. If there are several spot groups, these RV signals will add or cancel themselves, leading to a quasi-periodic signal with unpredictable but limited amplitude. Because the period of the RV signal varies between approximately 10 to 30 days, the idea is to average out these effects by an adequate sampling of the variation time scale.

In Paper I, a good observational strategy to average out oscillation and granulation noise was to take three 10-minute measurements per night with two hours of spacing. This strategy was applied to the actual calendar of HD69830, which is one of the most followed stars by HARPS. This calendar, besides observation gaps owing to bad weather, is very similar to 10 consecutive nights of observation every month. This idealistic strategy of three measurements per night of 10 minutes each, on 10 consecutive nights every month, will be referred to below as the 3N1 strategy. Because the typical time scales of activity noises are longer than 10 days (see Fig. \ref{fig:4}), the 3N1 strategy is not optimized to average out this kind of perturbation. In order to average out all three kinds of noise at the same time, we have to cover a period of a month with measurements, while keeping the benefit of the higher frequency sampling. Taking into account this last argument, we simulate the effect of two new observational strategies. Keeping the three measurements per night of 10 minutes and the 10 nights of observation per month, we change the separation between each observational night. The new considered observational strategy will measure the star every second night (hereafter 3N2 strategy) and every third night (hereafter 3N3 strategy). For the 3N1, 3N2, and 3N3 strategies the calendar of HD69830 was not suitable any more and we had to create new calendars, asking for the following criteria:
\begin{itemize}
\item The total observational time span is four years.
\item The star disappears from the sky during four months each year.
\item Each month of observation, the beginning of the measurements is randomly set. 
\end{itemize}
In addition, we randomly suppress 20\,\% of nights to take into account bad weather or technical problems. These calendars represent a total of 256 nights of measurements over a time span of four years.
\begin{figure*}
\begin{center}
\includegraphics[width=8cm]{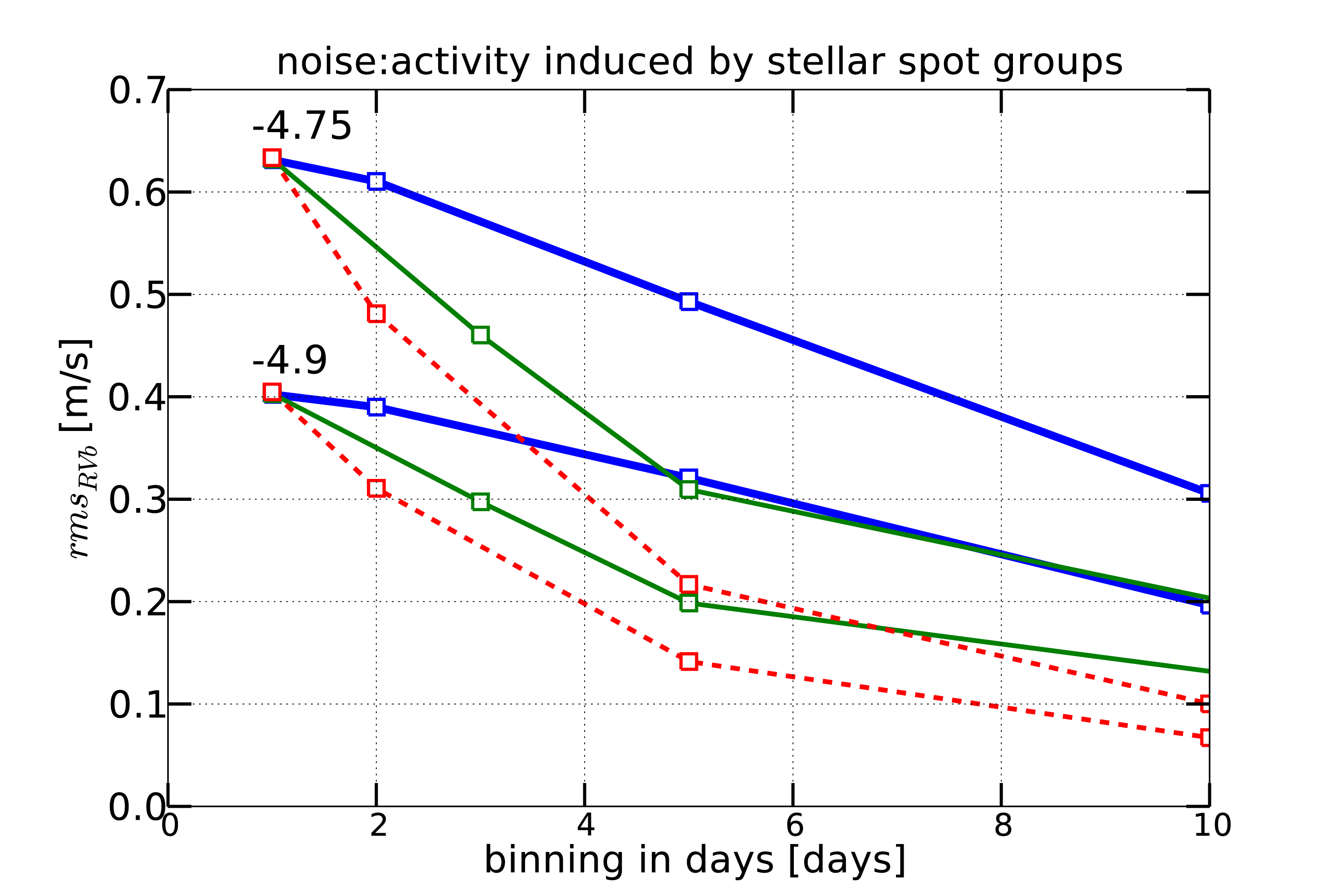}
\includegraphics[width=8cm]{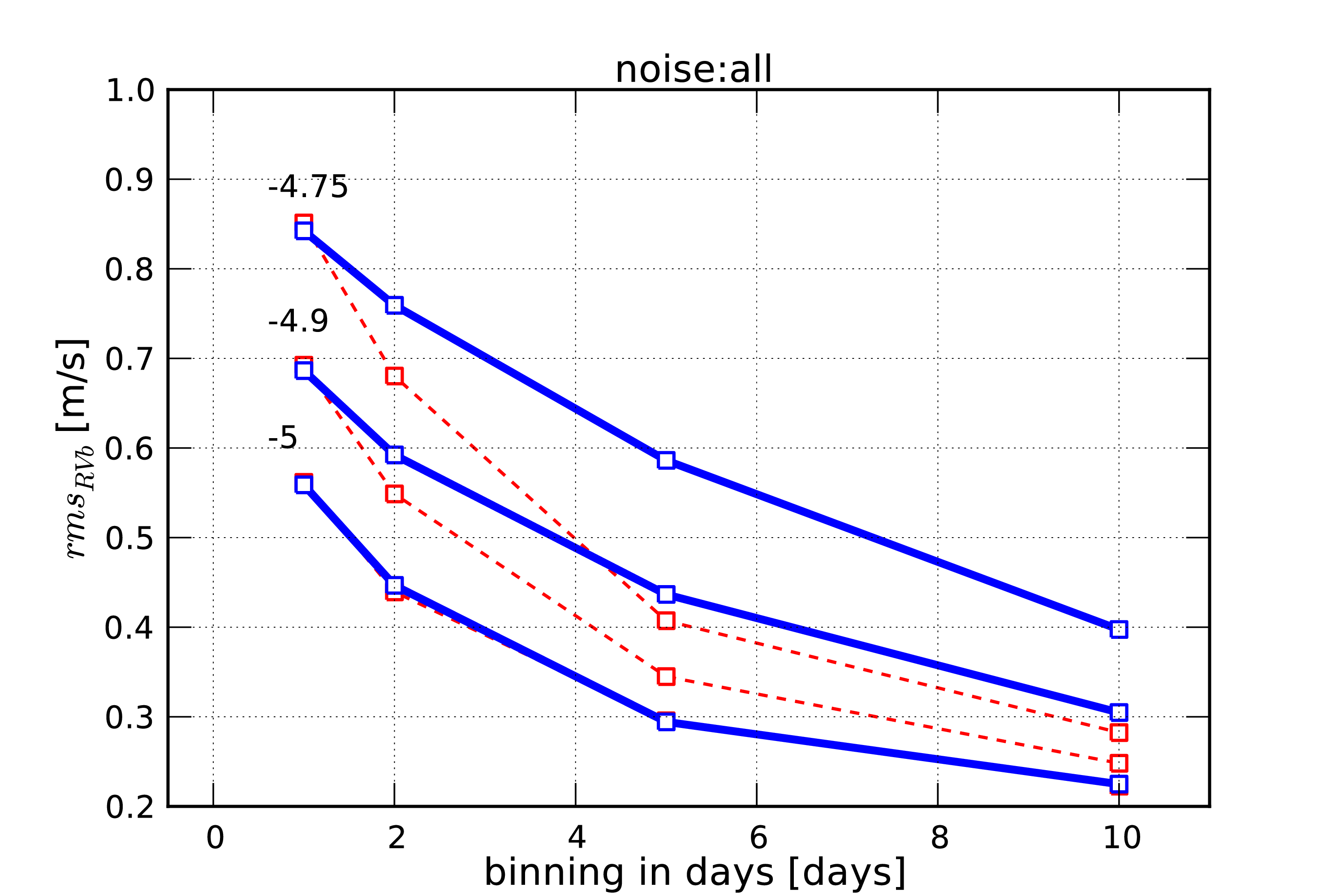}
\caption{Evolution of the binned RV variation, $rms_{RVb}$, owing to activity-related group spots (left panel) and to all kind of stellar noises (right panel) as a function of the binning in days. The three strategies shown for different $\log(R'_{HK})$ (given for each lines) use three measurements per night of 10 minutes and 10 nights of observation per month. The only difference lies in the spacing between each night of observation. Thicks lines correspond to the 3N1 strategy, thin lines to the 3N2 strategy, and dashed lines to the 3N3 strategy.}
\label{fig:5}
\end{center}
\end{figure*}
In Fig. \ref{fig:5} (left panel) we can compare these different observational strategies. The RV variation, induced by spot groups for a maximum level of activity ($\log(R'_{HK})=-4.75$), reaches a level of approximately 60\,cm\,s$^{-1}$. By using the 3N3 strategy, we manage to reduce this level of RV variation to 10\,cm\,s$^{-1}$ when averaging the effect in 10 day bins. For the same binning, the 3N1 strategy reduces this level to 30\,cm\,s$^{-1}$.

%For the 3 considered strategies, we take each time 3 measurements per night. This high frequency of measurement is essential to average out, after binning, the noises coming from oscillations and granulation (see Paper I). However, since the perturbation period of spot groups is longer than one night, this high frequency will not reduce the noise coming from activity. Thus, considering only activity noise, observing the star once or 3 times a night on 10 consecutive days every month will give a RV variation of 30\,cm\,s$^{-1}$, for a $\log(R'_{HK})$ of -4.75 and a binning of 10 days. In the last section of Paper I, we calculated the residual noise of HARPS measurement over 5 years of observation (obtained with only 1 measurement per night), without being able to attribute it to instrumental noise or activity noise. Since the residual noise after 10 days of binning is around 50 cm\,s$^{-1}$ and the activity noise around 30 cm\,s$^{-1}$ in the worst case, it seems that this residual noise is coming mainly from the instrument and as been identified to probably be due to the telescope guiding \citep[][]{Pepe-2008}.

We can now calculate $rms_{RVb}$ as a function of the binning in days, including all types of noise: oscillation, granulation, activity, and instrumental (80\,cm\,s$^{-1}$). The result for $\alpha\,Cen\,B\,(K1V)$ can be seen in Fig. \ref{fig:5} (right panel).

We notice that the 3N3 strategy averages out much more activity noise after binning than the 3N1 strategy.
%Taking measurements each 3 nights over 30 nights (3N3 strategy) averages out much more activity noise after binning then observing the star each night on 10 consecutive nights every month (3N1 strategy). 
Without activity ($\log(R'_{HK})=-5$) the two strategies are identical. The RV variations reachable using 10 days bins (suitable for long period planets) with the best considered strategy are 22\,cm\,s$^{-1}$, 25\,cm\,s$^{-1}$, and 28\,cm\,s$^{-1}$ for a $\log(R'_{HK})$ equal to -5, -4.9 and -4.75, respectively. If there is high activity, the improvement brought by this optimal strategy reaches 30\,\% compared to the 3N1 strategy. 

In conclusion, sampling the complete rotational period of the star with the same number of measurements, but with more separation between them, averages out the activity noise to a considerable extent. Thus, the strategy needs to be adjusted to the stellar rotational period. Here for the Sun, which has a 26 day rotational period, the 3N3 strategy is the optimal one with 10 measurements per month. For a star with a 20 day rotational period, the 3N2 strategy would average out the stellar noise better. In addition, increasing the gap between observational nights by more than three days for slow rotators is not recommended, because it will induce a poor sampling of short period planets.

%Notice that this precision is close to twice the RV effect induce by the Earth on the Sun, 9 cm\,s$^{-1}$.
%
\begin{figure*}[!t]
\begin{center}
\includegraphics[width=18cm]{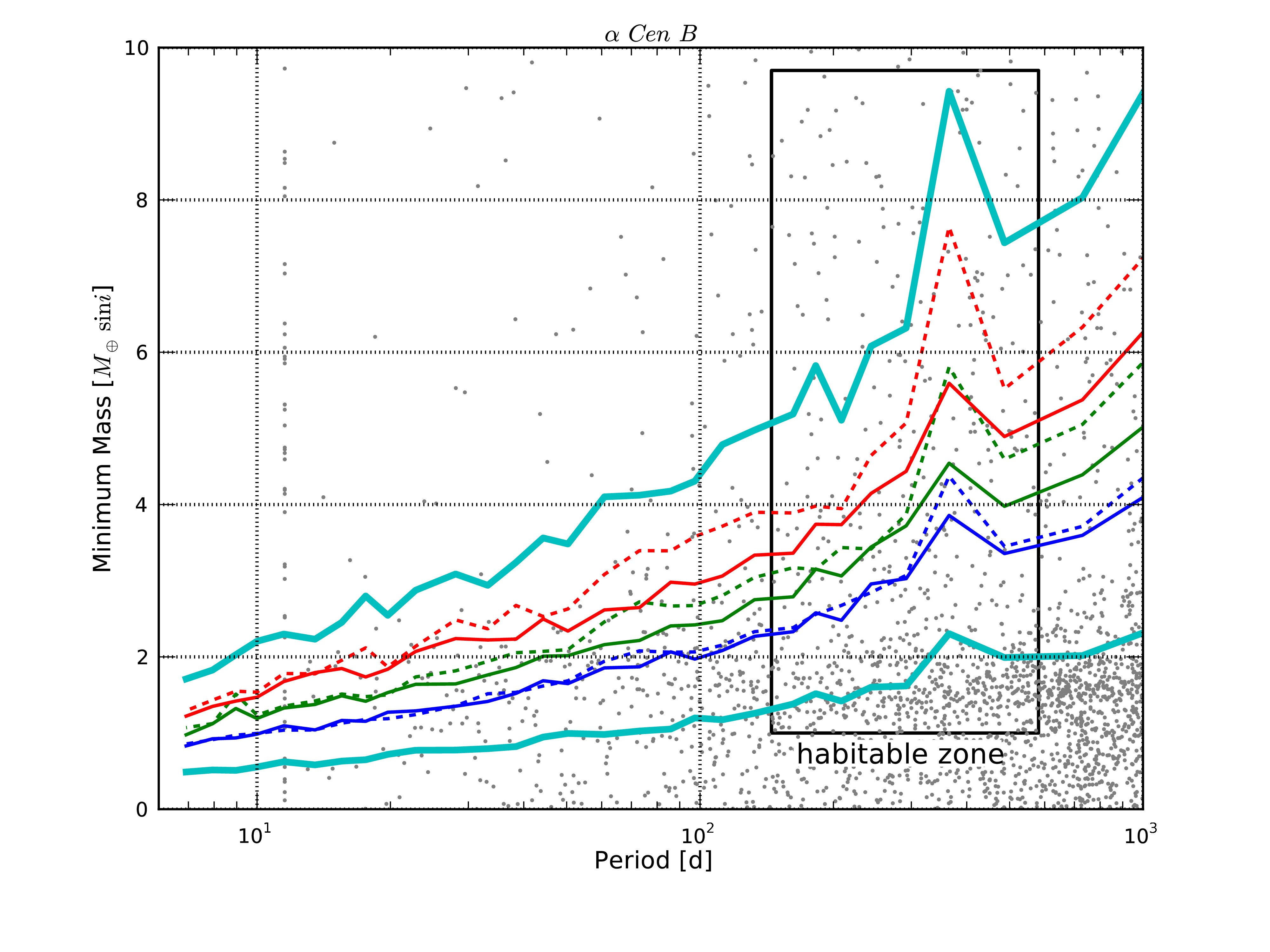}
\caption{Detection limits for $\alpha\,Cen\,B\,(K1V)$ for different activity levels and strategies. The thick line at the top corresponds to the present HARPS observational strategy (one measurement per night of 15 minutes, 10 consecutive days each month) for an activity level $\log(R'_{HK})=-4.75$. The second thick line at the very bottom corresponds to the detection limit, using ESPRESSO, for the 3N3 strategy and an activity level $\log(R'_{HK})=-5$. Besides these two thick lines, from top to bottom, the dashed thin lines correspond to the 3N1 strategy using HARPS, for an activity level $\log(R'_{HK})=-4.75$, $-4.9$ and $-5$, respectively. The 3N3 strategy using HARPS corresponds to the continuous thin lines on the same order of activity level. Finally, the small dots represent the planets expected from Bern's model with a random $\sin{i}$ for each body. We see that the cut-off in period around 12 days of the model introduces an anomaly in the planet distribution. The habitable zone is derived using the model of \citet{Selsis-2007} applied to the temperature and the luminosity of a K1V dwarf, calculated using the results of \citet{deJager-1987}.}
\label{fig:7}
\end{center}
\end{figure*}

\section{Detection limits in the mass-period diagram}\label{sec:3}

In order to derive detection limits in terms of planet mass and period accessible with the different studied strategies, we calculate the false-alarm probability (FAP) of detection using bootstrap randomization \citep[][]{Endl-2001,Efron-1998}. We first simulate a synthetic RV set containing stellar, instrumental, and photon noises as we did in the previous sections. The calendars used depend on the selected strategy, but all have a total of 256 observational nights over a time span of four years (see Sect. \ref{sec:2}). Then we carry out 1000 bootstrap randomizations of these RVs and calculate the corresponding periodograms. For each periodogram, we select the highest peak and construct a distribution of these 1000 highest peaks. The 1\,\% FAP corresponds to the power, which is only reached 1\,\% of the time.
%\footnote{In Paper I, we were calculating the FAP for each period. In the present paper, we calculate it for all the period at once, thus only one value is calculated for each bootstrap. Comparing the detection limit in the case of no activity with the 3N1 strategy (see Fig \ref{fig:7}) with the result of Paper I (see the case of $\alpha\,Cen\,B$ in Fig. 5 of the Paper I), we see that the 2 different techniques give exactly the same detection limits.}. 
The second step consists in adding a sinusoidal signal with a given period to the synthetic RVs. We calculate the periodogram of these new RVs and compare the height of the observed peak with the 1\,\% FAP. We then adjust the semi-amplitude of the signal until the power of the peak is equal to the 1\,\% FAP. The obtained semi-amplitude corresponds to the detection limit of a null eccentricity planet with a confidence level of 99\,\%. To be conservative, we test 10 different phases and select the highest semi-amplitude value.

In practice, we simulate 100 RV sets for each activity level ($\log(R'_{HK})$=-5, -4.9 and -4.75) and strategy, and to be conservative, only the 10 "worst" cases per strategy were considered. It is important to note that for each strategy we simulate the RVs according to the respective calendar (10 measurements per months on consecutive nights or every second or third night, eight months a year over four years) and we removed randomly 20 \,\% of the nights to simulate bad weather or technical problems. The observational calendars are therefore realistic. Figure \ref{fig:7} presents for different levels of activity the 99\,\% confidence level detection limits expected for the 3N1 and the 3N3 strategies using HARPS.

The thick line in top part of Fig. \ref{fig:7} corresponds to the detection limit for a maximum-activity level ($\log(R'_{HK})=-4.75$) using the present HARPS observational strategy (1N1 strategy = one measurement of 15 minutes on 10 consecutive days per month, see Paper I for details). Obviously, there is a great improvement brought by the 3N1 and 3N3 strategies compared to what is done at present on HARPS.

We can also compare the level of the detection limits obtained for the 3N1 and 3N3 strategies. For example, at a 100 day period, the improvement brought by the 3N3 strategy is 4\,\%, 10\,\% and 17\,\%, for an activity level of $\log(R'_{HK})=-5$, $-4.9$ and $-4.75$, respectively. As expected, the improvement increases with a rising activity level because the 3N3 strategy is optimized to reduce activity noise. For minimum activity, which corresponds to no spot groups and consequently no noise from activity in our simulation, the detection limits obtained with the two strategies should be identical. The 4\,\% difference can be explained by a more regular sampling (over all months with the 3N3 strategy and only 10 consecutive days per month for the 3N1). In conclusion, the 3N3 strategy seems to be an efficient strategy to average out all kinds of noise. For example, for a period of 200 days, which corresponds to the habitable region around a K1 dwarf such as $\alpha\,Cen\,B\,$, the detection limits for the 3N3 strategy ranges between 2.5 and 3.5 $M_{\oplus}$ depending on the activity level. Following these results, it seems clear that even including the activity noise related to spot groups, planets smaller than 5 $M_{\oplus}$ in habitable regions could be detected with HARPS using an appropriate observational strategy.

Future instruments, like ESPRESSO at the VLT, will reach a precision level of 10\,cm\,s$^{-1}$. The thick line at the very bottom of Fig. \ref{fig:7} shows the improvement when we 
change the instrumental noise from 80 to 10\,cm\,s$^{-1}$. Evidently the improvement is significant, reducing approximately the mass detection limit by 1 $M_{\oplus}$. We note, however, that because we used HARPS data as a basis for our simulations, it is possible that we overestimate the noise of the ESPRESSO-like radial-velocities. These latter values must then be seen as upper limits.

\section{Planet detection according to Bern's model}\label{sec:4}

The detection limits give us an idea of the smallest planet that could be found with an appropriate strategy. Nevertheless, this does not inform us on the proportion of small mass planets that could be detected. This requires a model of planetary formation that can predict the "true" population of small planets. We used Bern's model \citep[e.g. ][]{Mordasini-2009a,Mordasini-2009b}, which is a good proxy of planet population. In their simulation they solve as in classical core accretion models the internal structure equation for the forming giant planet, but include at the same time disk evolution and type I and II planetary migration.

\begin{figure*}
\begin{center}
\includegraphics[width=8cm]{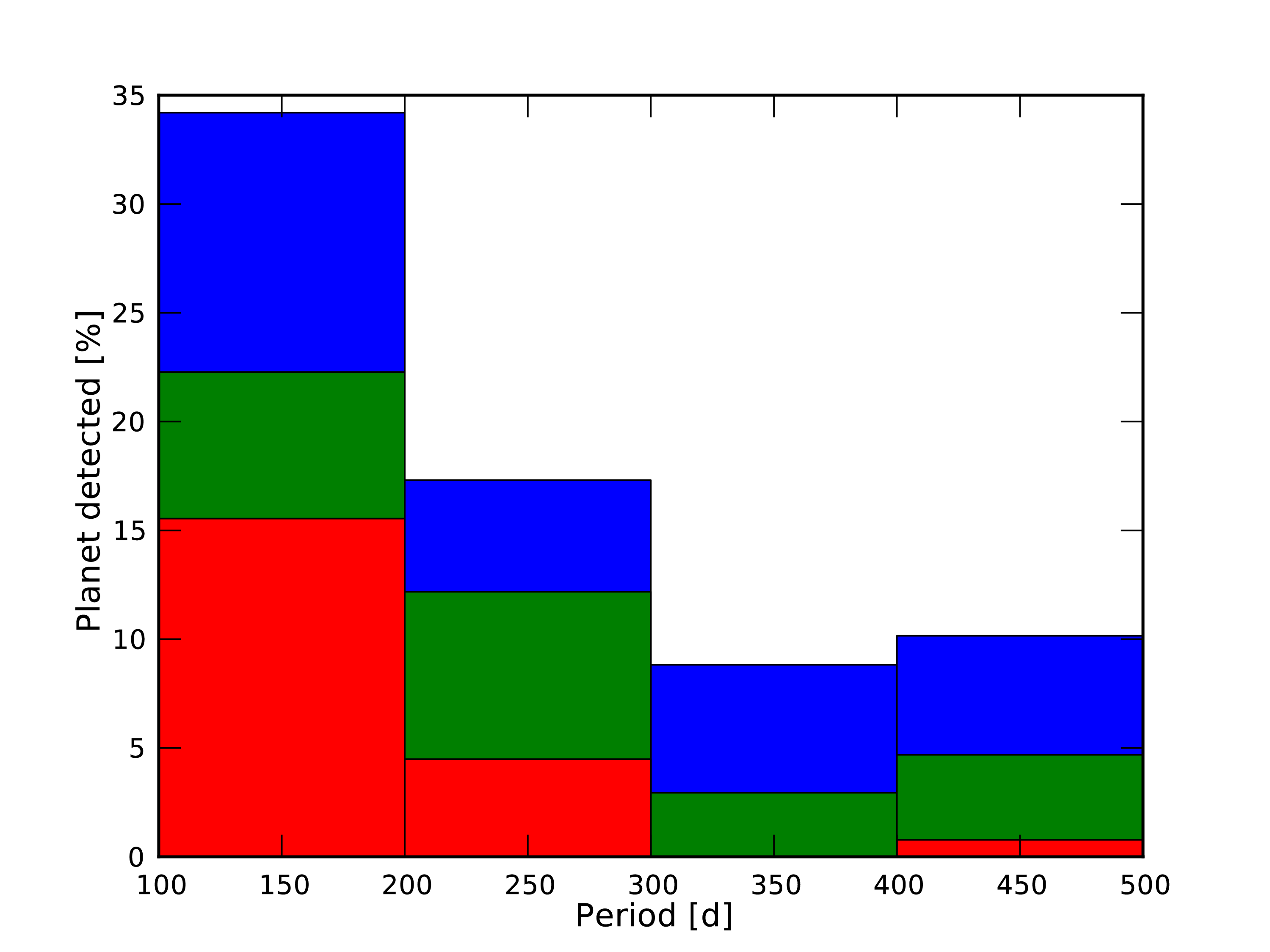}
\includegraphics[width=8cm]{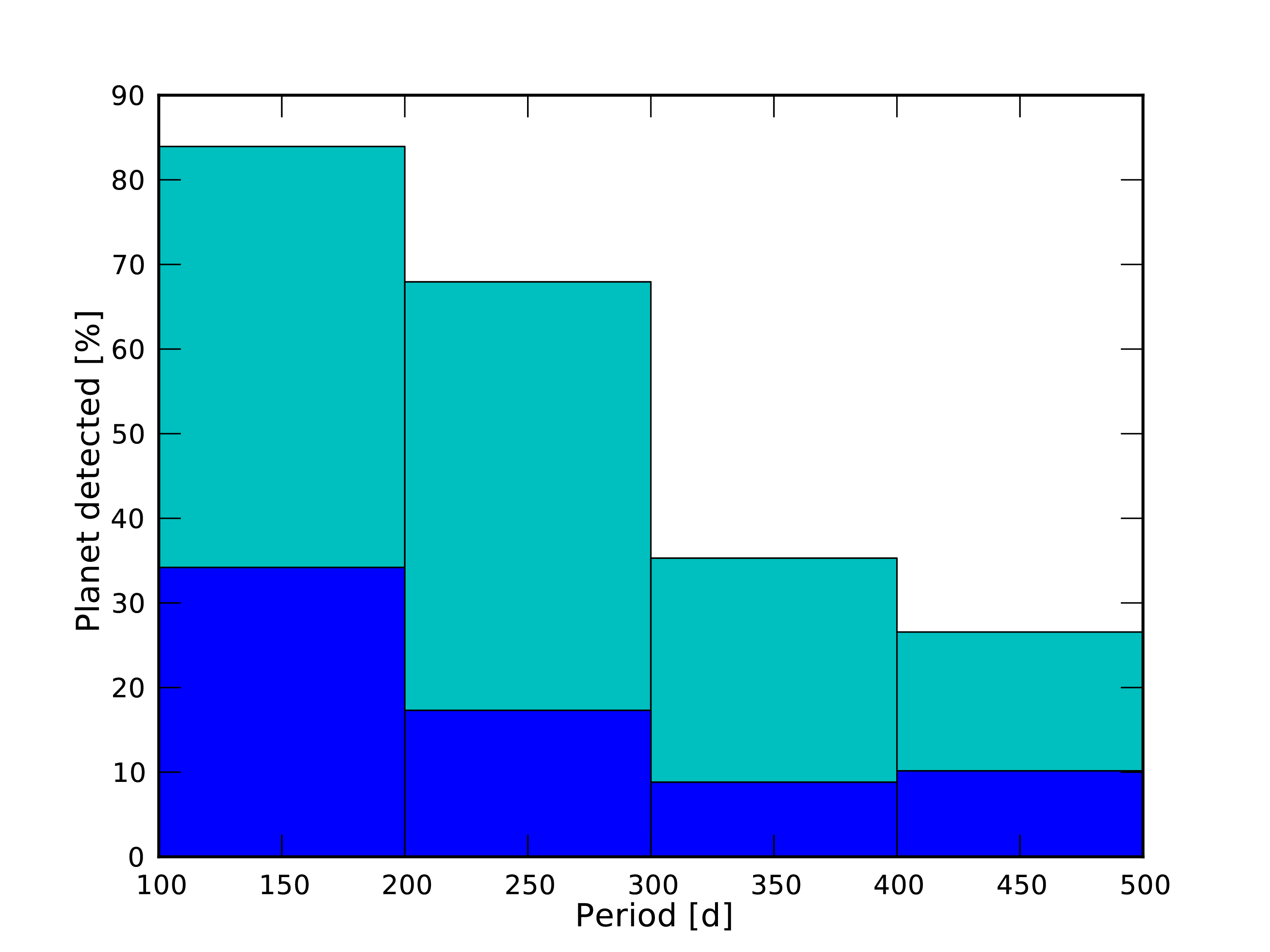}
\caption{\emph{Left panel:} Proportion of expected planets with 1 $\le M\sin{i} \le$ 5 $M_{\oplus}$ that could be found with HARPS around early-K dwarfs, using the 3N3 strategy. For each bin from top to bottom we have the detection proportion for $\log(R'_{HK})$ equal to $-5$, $-4.9$ and $-4.75$. \emph{Right panel:} Proportion of expected planets with 1 $\le M\sin{i} \le$ 5 $M_{\oplus}$ that could be found with HARPS (dark or blue) and ESPRESSO (gray or cyan) around early-K dwarfs, using the 3N3 strategy. The activity level is set to a $\log(R'_{HK})$ equal to $-5$.}
\label{fig:8}
\end{center}
\end{figure*}

Using the detection limits derived in Sect. \ref{sec:3}, we calculate the number of planets predicted by Bern's model that are above a given detectable level. The proportion of expected planets between $1$ and $5\,M_{\oplus}$ that could be found with the 3N3 strategy on HARPS is represented in the left panel of Fig. \ref{fig:8}. In this panel, we clearly see that the proportion of detection decreases when the activity level increases. For the period range between 100 and 200 days, the proportion of planets with $1M_{\oplus}<M_{\oplus}\sin{i}<5M_{\oplus}$ that would be found with HARPS changes from 35\,\% to 15\,\% when the $\log(R'_{HK})$ varies from $-5$ to $-4.75$. Although the activity greatly reduces the number of detectable planets, it is still possible to find some in the habitable region of K dwarfs (200 days) for the highest case of activity ($\log(R'_{HK})=-4.75$ for \emph{low-activity} stars, see Sect. \ref{sec:1}). 

We note for a similar activity level that the value of the bin for 300 to 400 days is lower than the one for 400 to 500 days. Because the mass detection limits increase with period, we expect the opposite. This effect is owing to the observational calendar used. Indeed, because stars in the sky can generally be followed during eight months per year before they disappear (an effect taken into account in our observational calendar), the one year period is not well sampled, which complicates the detection. This is well illustrated in Fig. \ref{fig:7}, were a peak near one year appears for each detection limit.

A comparison with what could be found using ESPRESSO is shown in the right panel of Fig. \ref{fig:8}. If the activity level is set to $\log(R'_{HK})=-5$, ESPRESSO could find 80\,\% of the planets in a mass range from 1 to 5\,$M_{\oplus}$ and a period between 100 and 200 days, whereas HARPS could only find 35\,\% of them. 

\section{Detection of a simulated planet}

To show that the calculated detection limits are realistic, we present in this section an example of planet detection. For $\alpha\,Cen\,B\,(K1V)$, we first create some synthetic RVs including all types of noise, oscillations, granulation phenomena, and activity. The strategy used is the 3N3 and the activity level is set to $\log(R'_{HK})=-4.9$. Secondly, we add the RV signal induced by a small planet in the habitable zone ($M\sin{i} = 2.5\,M_{\oplus}$, a 208 day period). To check whether the planet is detected or not, we calculate the periodogram and the 1\,\% and 0.1\,\% FAP, using bootstrap randomization (see Sect. \ref{sec:3}). 

\begin{figure*}
\begin{center}
\includegraphics[width=6cm]{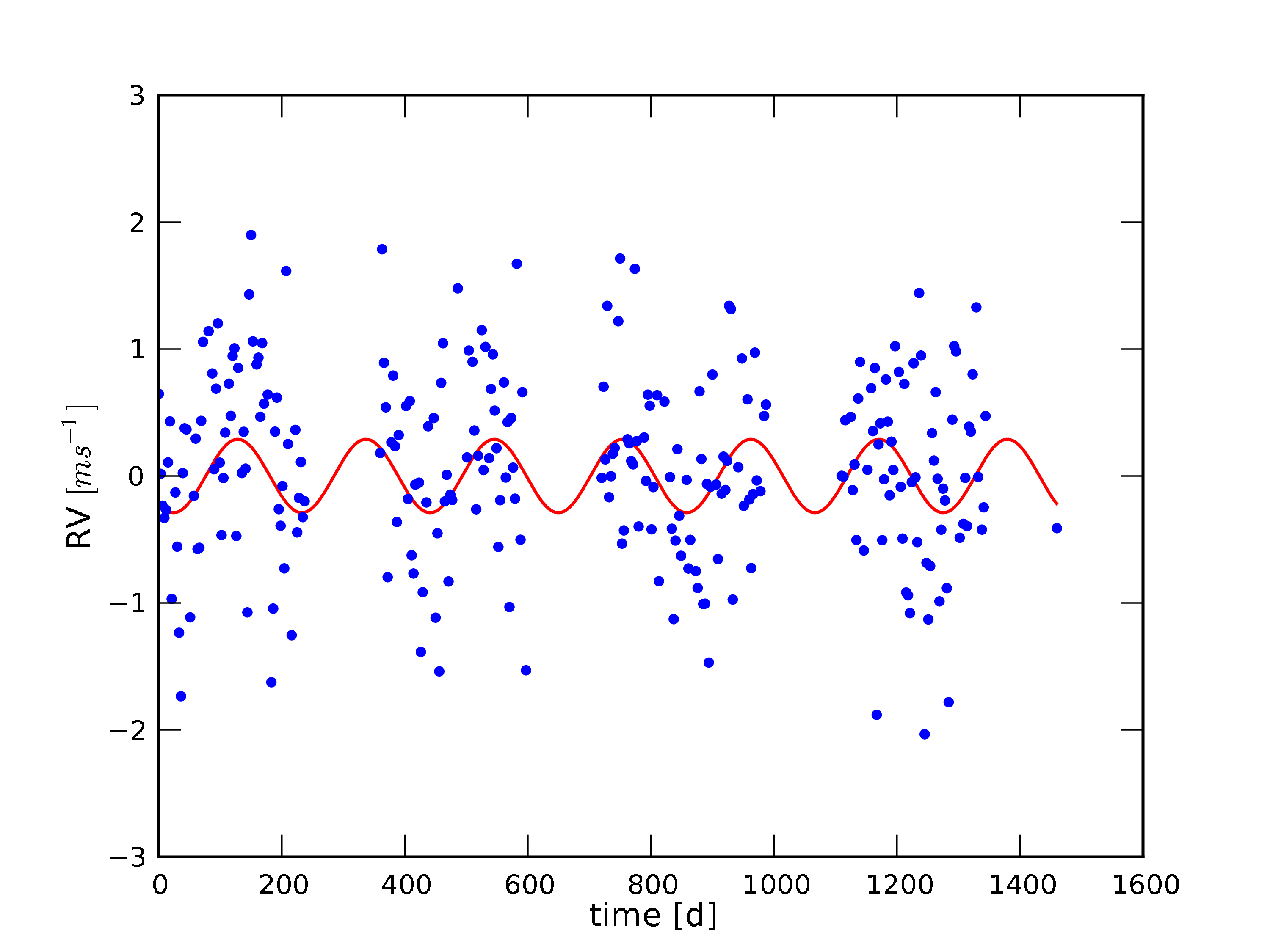}
\includegraphics[width=6cm]{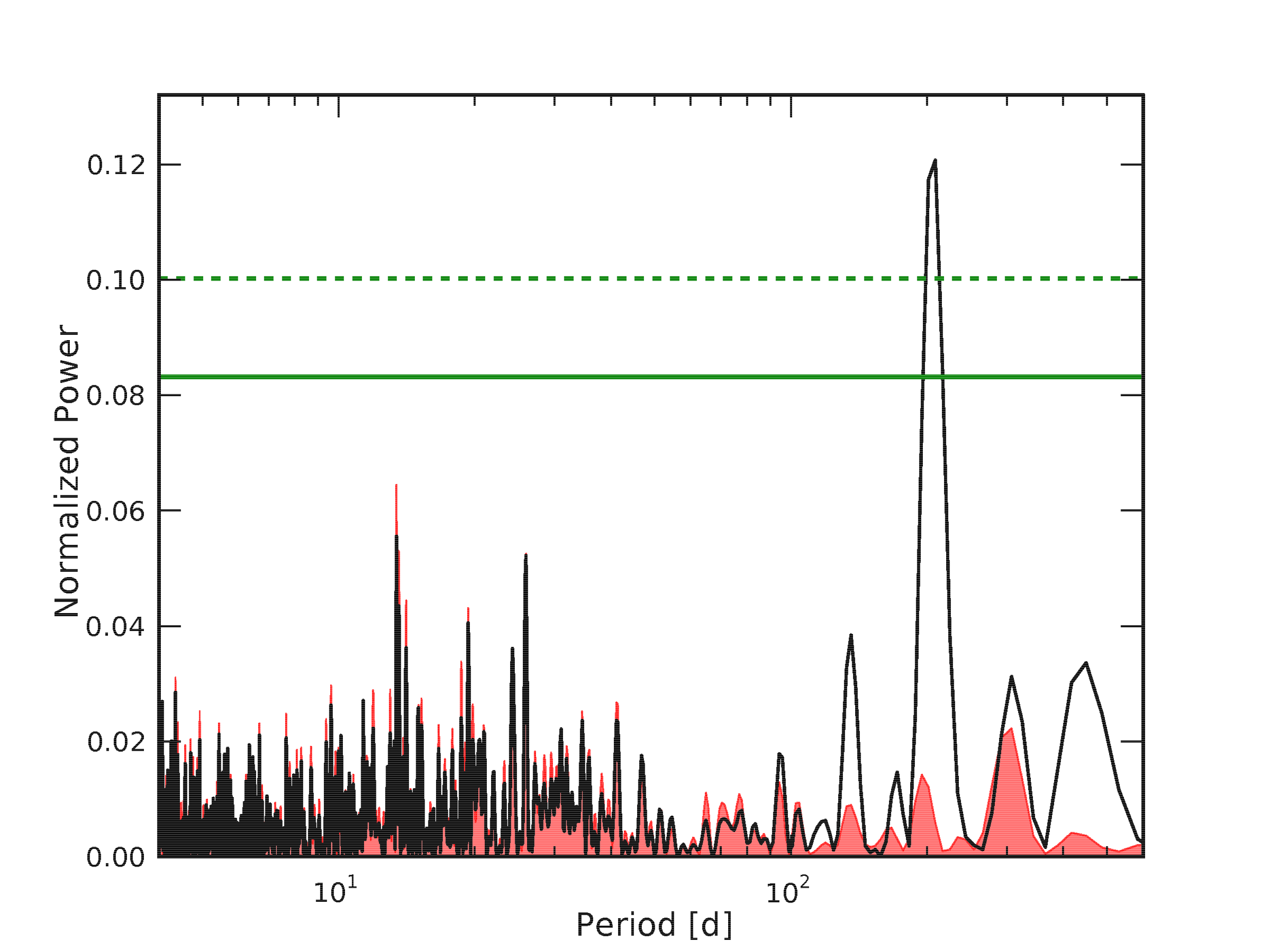}
\includegraphics[width=6cm]{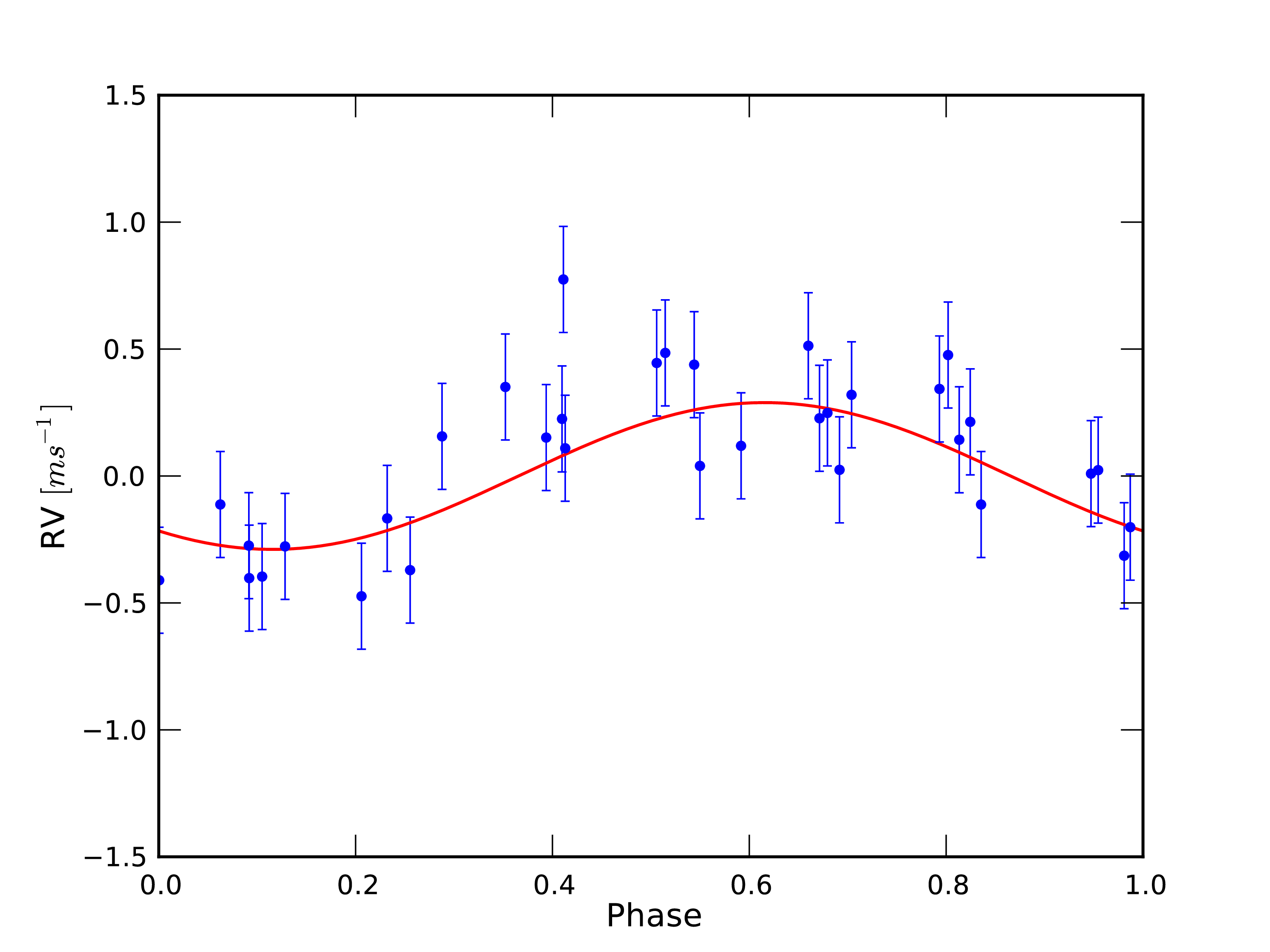}
\includegraphics[width=6cm]{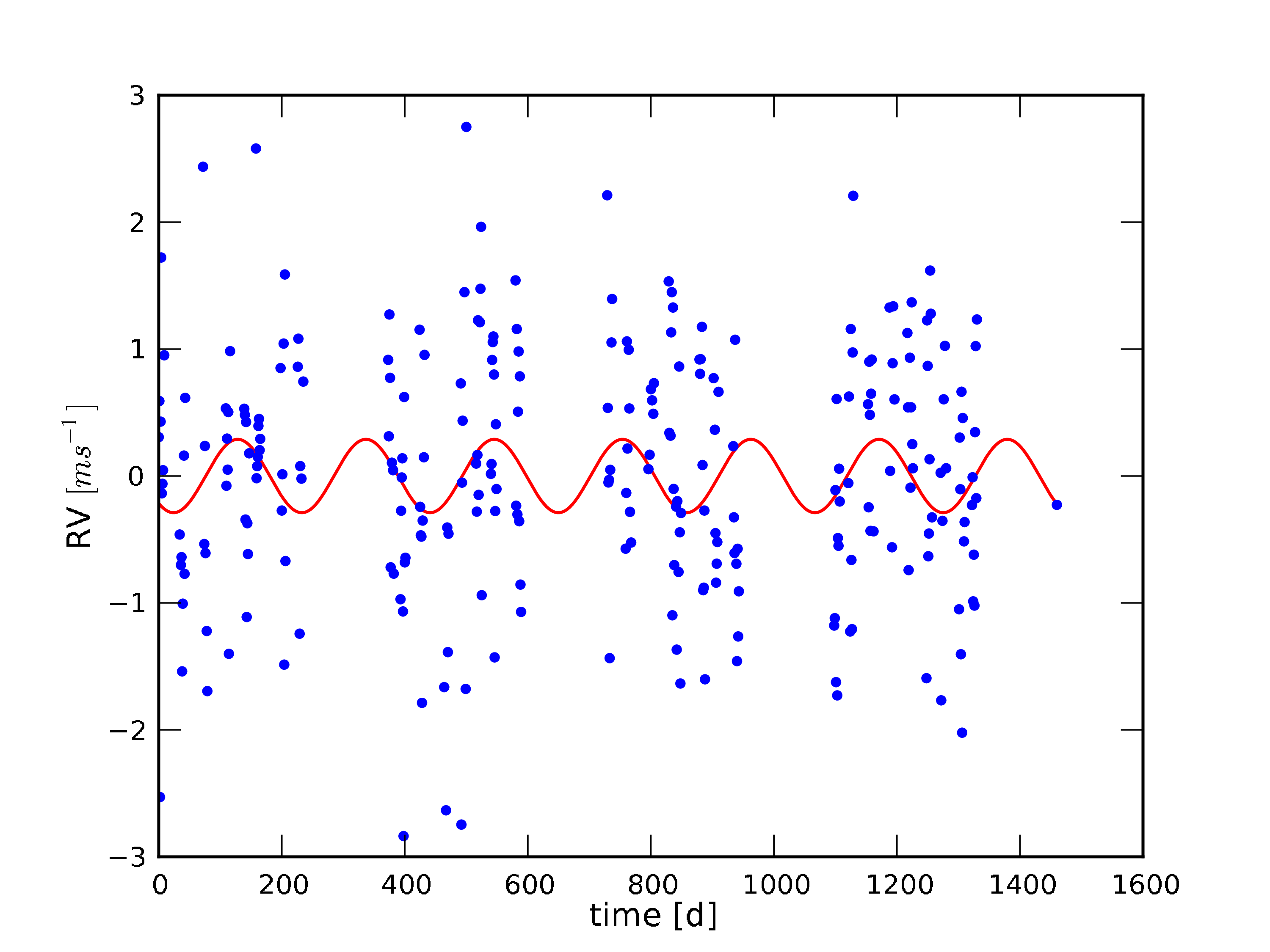}
\includegraphics[width=6cm]{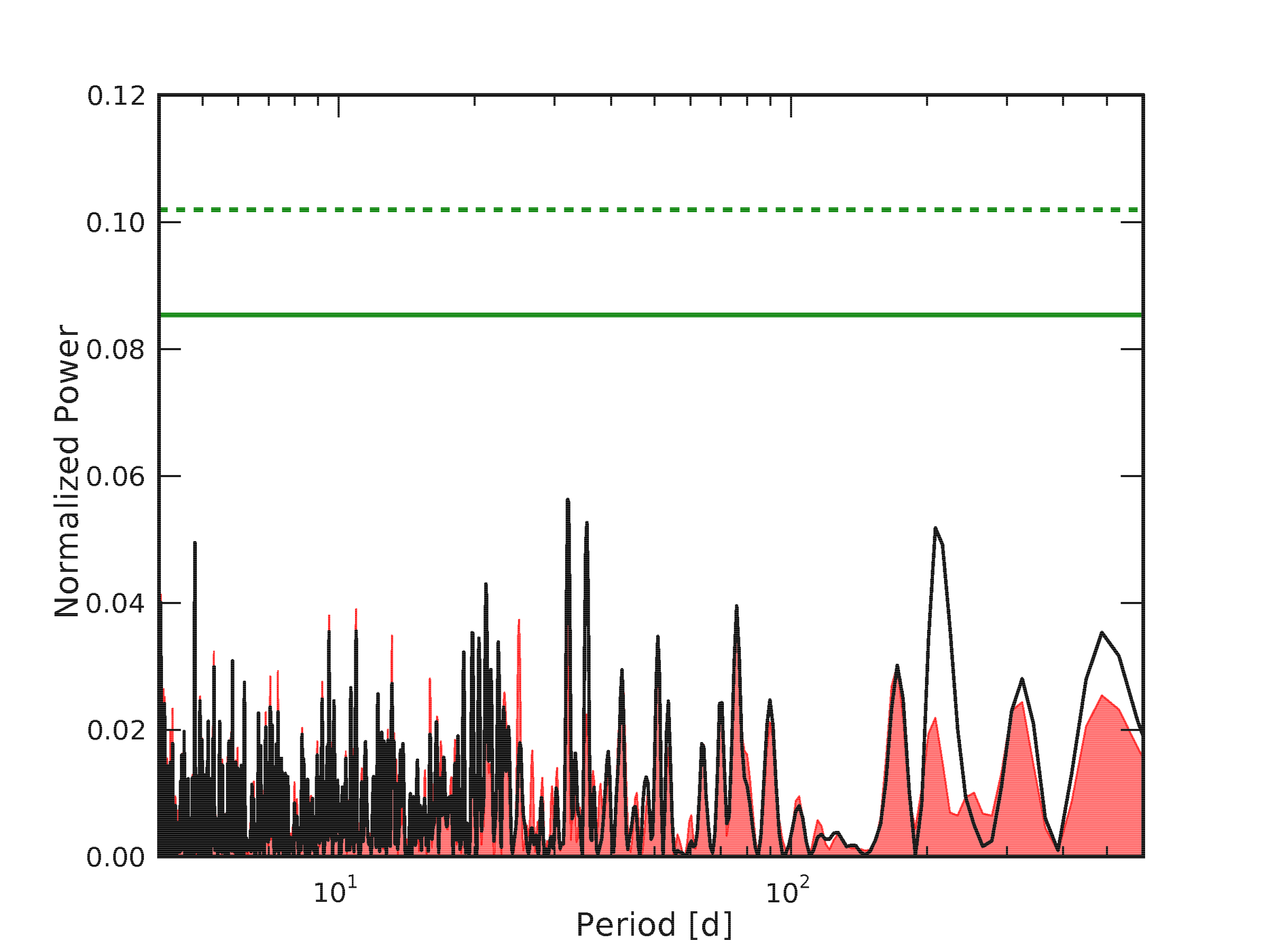}
\includegraphics[width=6cm]{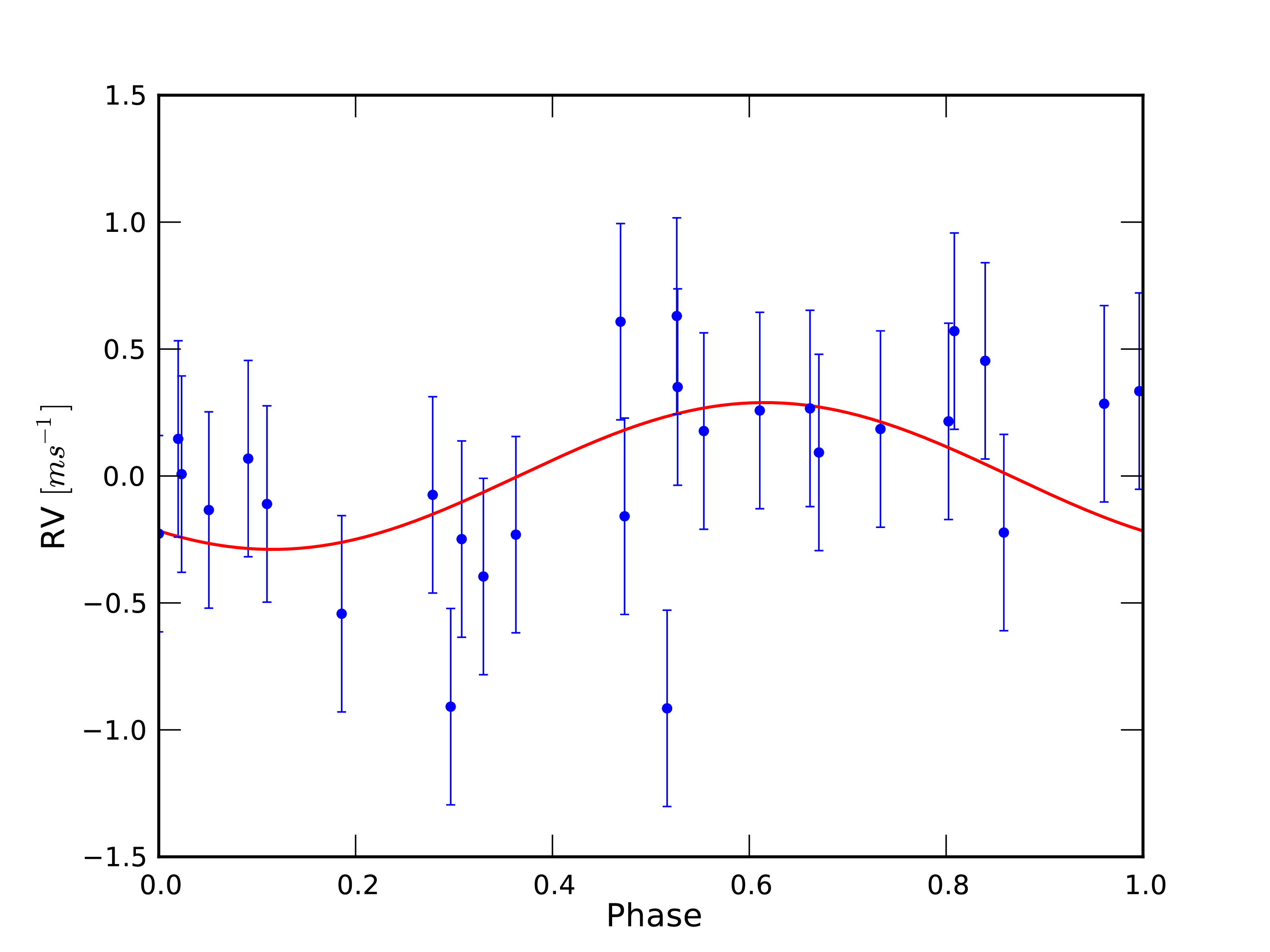}
\caption{Detection simulation for a $2.5\,M_{\oplus}$ planet in the habitable zone ($P=208$ days) of a K1 dwarf presenting an activity level $\log(R'_{HK})=-4.9$. \emph{Top panels:} Detection simulation for the 3N3 strategy. On the left, we show the raw RVs including the stellar noise (oscillation, granulation phenomena, and activity) plus the injected planet (red curve). In the middle, the periodogram of the stellar noise only (red filled curve) and the stellar noise plus planet (black line) is represented. The horizontal lines correspond from top to bottom to the 0,1\,\% and the 1\,\% FAP. Finally we show on the right the raw RVs binned on 1 month and folded in phase with the period of the planet. \emph{Bottom panels:} The same but for the 1N1 strategy (strategy presently used for the high-precision HARPS program, one measurement per night of 15 minutes on 10 consecutive days per month).}
\label{fig:9}
\end{center}
\end{figure*}

In Fig. \ref{fig:9}, top panels, we show the raw RVs, stellar noise plus planet, for the 3N3 strategy as well as the added planet. The periodogram with the 1\,\% FAP for each period is also represented. The peak at 200 days is far above the 0.1\,\% FAP, which proves that the planet is easily detected. To compare this with the actual measurements made on HARPS for the high-precision program, we apply the same process for the 1N1 strategy (1N strategy in Paper I). This strategy consists of measuring the star 15 minutes per night on 10 consecutive days per month. The bottom panels of Fig. \ref{fig:9} show the results. Evidently here the planet is far from being detectable.

Looking at Fig. \ref{fig:7}, we see that the mass detection limit for 208 days of period and for an activity level $\log(R'_{HK})=-4.9$ is more than 3\,$M_{\oplus}$ for the 3N3 strategy. In this example, we see that using the same strategy and configuration, a 2.5 earth mass planet could easily be detected. Thus the detection limits derived in Sec. \ref{sec:3} are very conservative. This is because we consider only the 10 "worst" cases when deriving the detection limits.

\section{Concluding remarks}

In Paper I we had proposed an efficient observational strategy to reduce the stellar noise generated by oscillation and granulation as much as possible. This strategy, requiring three measurements per night of 10 minutes over 10 consecutive days each month, improves the averaging out of stellar noise coming from oscillation and granulation by 30\,\%, with an observational cost only multiplied by a factor 2. In the present paper, we add the noise induced by activity-related stellar spot groups.

To study the radial-velocity (RV) effect caused by stellar spot groups, we consider three different activity levels. Based on Sun observations, we simulate the minimum solar activity level ($\log(R'_{HK})=-5$), the maximum one ($\log(R'_{HK})=-4.75$), and an intermediate one ($\log(R'_{HK})=-4.9$). Comparing the RV variation at maximum activity given by our simulation with the one calculated by real position and size of spot groups from cycle 23 \citep{Meunier-2010a}, we find a very good agreement, 51\,cm\,s$^{-1}$ and 48\,cm\,s$^{-1}$, respectively. We note that the RV effect of activity is not fully simulated because the inhibition of convection in active regions is not yet implemented. According to \citet{Meunier-2010a}, this effect could be important and consequently, the detection limits calculated here could be underestimated. Compared to the results of \citet{Meunier-2010a}, our simulation is more general in the sense that we can predict the RV effect of activity related spots groups for other stars, knowing a few characteristics (mean spot number, distribution of spot groups lifetime, presence of active longitudes, differential rotation). This will be very important when a better knowledge of activity phenomena of other stars than the Sun will be known. The Kepler mission should give us some clues in the near future.

Modeling the short-term activity with a three sine waves function with period $P_{fit}$, $P_{fit}/2$ and $P_{fit}/3$ can greatly reduce the spot-induced noise by approximately 70\,\%. However, we have to be very careful with this method because $P_{fit}$ can vary in a non negligible way from the rotational period of the star. Thus, signal of small mass planets with period similar to the rotational period of the star will be killed by this type of model. Only a study of other CCF parameters such as the FWHM or the BIS can give us clues on the true nature of the signal: short-term activity or planet.

After simulating the effect of several observational strategies, the most efficient one to average out all kinds of noise is the 3N3. This strategy consists in measuring the star three times 10 minutes per night, with a spacing of two hours. Then the measurements are taken every third night, 10 days every month. With this strategy it would be possible to find planets of 2.5 to 3.5 Earth mass with HARPS in the habitable region of K dwarfs (200 days). The first mass value corresponds to a case without activity ($\log(R'_{HK})=-5$) and the second one to a maximum activity level ($\log(R'_{HK})=-4.75$). Even if the activity caused by spot groups introduces a non negligible noise, small mass planets in habitable regions could be detected with HARPS with an appropriate observational strategy.

%To take into account this activity effect, based on Sun observations, we simulate the evolution of spot groups on the surface of a rotating star. Since the number of spot groups change during the 11-year activity cycle, we made 2 simulations, one for a $\log(R'_{HK})$ equal to $-4.9$ and another one for a $\log(R'_{HK})$ equal to $-4.75$. The RV effect induced by these spots was calculated and we found that the period of activity perturbations is between 10 and 30 days. Since this period is larger than the perturbation period for oscillation and granulation, the best strategy derived in Paper I is no more suitable when the activity rise up. Keeping the 3 measurements per night of 10 minutes strategy to average out oscillation and granulation, we found that the best strategy, with an equal total observation time, was to observe the star each 3 nights, 10 days a month. This new strategy manage to lower the detection limits up to 50\,\% for long periods.

%The presence of activity is a problem for planets presenting a period similar to the typical time scales of activity effects, 10 to 50 days. In the case of shorter period planets, the RV signal will not be disturb by activity perturbations and in the case of longer period planets, we can use a huge binning, which will average out activity effect without reducing the RV amplitude induced by the planets. It is thus easier to average out activity effects for long period planets than for intermediate period ones (10 to 50 days)

Using the population of low mass planets predicted by Bern's model, we calculate the proportion of planets that could be found with the 3N3 strategy. Using HARPS, it would be possible to find 35\,\% of the planets below $5\,M_{\oplus}$ with a period between 100 and 200 days and a low-activity level ($\log(R'_{HK})=-5$). This value decreases to 15\,\% for a $\log(R'_{HK})$ equal to $-4.75$. If we trust this model of formation, which is the most compatible with observations, HARPS could discover several planets below $5\,M_{\oplus}$ in the habitable region of early-K dwarfs.

In our simulation, the 3N3 strategy appears to be the best one to average out activity noise. This is because of the rotational period of the Sun, which is fixed in our simulation at 26 days. Indeed, when choosing 10 observational nights a month, the best way to sample the entire rotational period is to observe the star every third night. For shorter rotational periods, increasing the measurement frequency would lead to a better averaging. For longer rotational periods, reducing the measurement frequency is not recommended because short period planets will be poorly sampled.

Taking three measurements of 10 minutes per night gives us a very low photon noise when binning the data over the night. Thus, the photon-noise is not a limitation any longer, and 2 to 4-meters class telescopes can still be used to go down to the 10\,cm\,s$^{-1}$ level on bright stars. However, spectrographs must also intrinsically reach this level of precision and for the moment, only ESPRESSO@VLT (http://espresso.astro.up.pt/) is designed to reach this goal. With such a precision, improvements would be remarkable. The detection limits would go down in mass by a level of 1 $M_{\oplus}$, reaching 1.3 $M_{\oplus}$ in the habitable region of early-K dwarfs. In addition, ESPRESSO would find 80\,\% of the planets between 1 and 5 $\,M_{\oplus}$ and with a period between 100 and 200 days, which is more than twice as much as what HARPS could find. Moreover, CODEX@E-ELT \citep[e.g][]{Pasquini-2008}, would improve the long term instrumental precision down to a few cm\,s$^{-1}$ level and would detect $1\,M_{\oplus}$ planets in habitable regions.

\begin{acknowledgements}

X. Dumusque would like to acknowledge Y. Alibert, J. Hall and W. Lockwood for private communication. We acknowledge the support by the European Research
Council/European Community under the FP7 through Starting Grant agreement number 239953. NCS also acknowledges
the support from Funda\c c\~ao para a Ci\^encia e a Tecnologia (FCT) through program Ci\^encia 2007 funded by FCT/MCTES (Portugal) and POPH/FSE (EC), and in the
form of grants reference PTDC/CTE-AST/098528/2008 and PTDC/CTE-AST/098604/2008. SGS is supported by grant SFRH/BPD/47611/2008 from FCT/MCTES.. Finally, this work was supported by the European Helio- and Asteroseismology Network (HELAS), a major international collaboration funded by the European Commission's Sixth Framework Program (grant : FP6-2004-Infrastructures-5-026183).

\end{acknowledgements}

\bibliographystyle{aa}
\bibliography{dumusque_bibliography}

\end{document}